# Geometrically exact static 3D Cosserat rods problem solved using a shooting method

*SURMONT Florian*[a,*], *COACHE Damien*[a]

[a] *Bureau Veritas, 8 avenue Jacques Cartier, 44800 Saint-Herblain, France*

## Abstract

This paper aims to solve the equations of geometrically exact 3D nonlinear Cosserat static rods under large displacements with a mesh free alternative method to finite elements: the shooting method. One of the main goal of the work presented is to introduce a new multiple shooting method to handle geometrical discontinuities as well as beams assembled through mechanical linkages in a multi-body framework. An additional purpose is to propose a new shooting method algorithm in which the dual set of Modified Rodrigues Parameters (MRP) and Shadow Modified Rodrigues Parameters (SMRP) are used for rotations parametrization along the length of the beam in order to model extremely large rotations. Several numerical examples demonstrate the validity, the performance and accuracy of the proposed approach.

[*] Corresponding author

Email address: florian.surmont@bureauveritas.com (SURMONT Florian)

## 1. INTRODUCTION

The study of beams assemblies undergoing large displacements with nonlinear couplings remains a strong topic of interest with applications in various fields such as civil engineering, aeronautics or marine renewable energies in which the use of discretized methods such as finite element is largely dominant. This paper aims to solve geometrically exact 3D nonlinear Cosserat static rods by means of a shooting method [1]. The main benefit of shooting methods over finite difference or finite element methods is that they naturally and automatically transfer the problems of error control and stepsize variation to the initial value solver, relieving the user of these difficult and critical tasks [2].

Shooting methods are not widely used for solving mechanisms involving flexible beam assemblies and main contributions come from robotics fields [3] and medical fields [4] for their quasi-analytical accuracy and their computation time efficiency [5]. However the literature focuses on solving nonlinear uniform beams only with single shooting or problem specific multiple shooting methods. The single shooting method shows its limits as soon as material discontinuities or beams assemblies are involved because specific processing must be made at discontinuities interfaces depending on the problem to solve. Thus no generic formulation can be found in the literature. Moreover rotation parametrization is a key point in the modeling of largely flexible rods. Literature either considers rotation matrices directly integrated and constraints generated from the rotation vector [6] or parametrized with successive angles of revolute joints mechanism [3]. Such parametrizations are not satisfactory either numerically or within a static multi-body framework. Indeed, numerical integration as well as numerical inversion can lead respectively to accumulated errors and convergence problems when studying largely deflected beams.

This paper proposes a solution for both problems: an efficient rotation parametrization of beams equations based on Modified Rodrigues Parameters (MRP) and a new multiple shooting method which is the combination of the classical multiple shooting method and a multi-body approach.

This work is the extension of previous authors' work on the use of the shooting method to model 3D quasi-static mooring lines and cables behavior as string elements [7]. Like strings, static rods are governed by a two-point boundary value problem (TPBVP). The shooting method is precisely a dedicated tool to solve such problem [8].

In this paper, the static 3D TPBVP of rod equations is set up with the introduction of arbitrary 3D external distributed loads following the Cosserat formalism developed in [9]. Apart from the Timoshenko assumption stating that every cross section behaves as rigid bodies, no other assumptions are made, especially no small displacement nor deformation are considered and geometrical deformation are kept fully nonlinear. The ordinary differential equations (ODEs) are kept continuous, meaning that no discretization of the geometrical or internal load fields is performed.

The modified Rodrigues parameters (MRP) are chosen to parameterize the rotation matrix as they constitute a minimum set of three degrees of freedom and have the advantage of avoiding the well-known gimbal lock kinematic singularity of the widely used Euler angles. Rods kinematic and balance equations are then written with respect to MRP.

The resultant contact forces and moments are seen as generalized internal forces. A constitutive law is used to connect generalized strains to generalized internal forces.

By means of the multibody analysis perspective, a dedicated representation of mechanical linkages through boundary conditions is proposed, with special emphasis on their modeling into kinematic joints and their use in association with the shooting method.

The TPBVP is then solved iteratively as a succession of initial value problems (IVPs) using the shooting method. The single and multiple shooting methods principles and applications to the current problem are reminded.

Finally, in order to validate the approach and the accuracy of the single shooting method, five validation cases based on the classical rod mechanics literature and four more based on material strength are investigated.

## 2. TPBVP EQUATION

The TPBVP for static rods consists in a set of ordinary differential equations (ODEs) associated with two boundary conditions at segment extremities. Cosserat rod theory is derived from Antman [9] formalism. The Cosserat, Timoshenko or Simo & Riflex are based on the same assumption: every cross section behaves as a rigid body. No assumption on small deformations and displacements is made so the rod can undergo large deformations in space suffering flexure, torsion, extension and shear. Finally rod's assumptions can be summarized as follow:

(i) the configuration is defined as a set of oriented material points (sections) having the geometrical property to be a curve in $\mathbb{R}^3$,

(ii) every cross section behaves as a rigid body

(iii) geometrical and mechanical functions are sufficiently regular to be derived accordingly

## *2.1 Geometrical Deformations*

From (i), a material point in space is described by a global vector geometrical field $\boldsymbol{r}(s)$ and two vectors $\boldsymbol{d_1}$ and $\boldsymbol{d_2}$ lying in the section as shown in Figure 1. Assumption (ii) imposes that the angle between $\boldsymbol{d_1}$ and $\boldsymbol{d_2}$ remains constant. For convenience $\boldsymbol{d_1}$ and $\boldsymbol{d_2}$ are set orthogonal. We can then define another vector $\boldsymbol{d_3}$ orthogonal to the two pre-cited such that $\boldsymbol{d_3} = \boldsymbol{d_1} \times \boldsymbol{d_2}$. These three vectors are called directors and recast into the second geometrical field: rotation matrix $R(s)$ representing the section orientation. Both geometrical fields depend on a parameter $s$ identified as the curvilinear abscissa of the reference configuration (unconstrained configuration) $\left(\boldsymbol{r}^0(s), R^0(s)\right)$ such that $\partial_s \boldsymbol{r}^0(s) = \boldsymbol{v^0}(s)$ and $\partial_s R^0(s) = \tilde{u}^0(s) R^0(s)$, where $\partial_s \cdot$ denotes the partial derivation with regards to parameter $s$. On the reference configuration, the curvilinear abscissa is defined on the interval $s \in [0, L]$. The unstretched length of the material segment is $L = \int_0^L \left|\boldsymbol{v^0}(s)\right| ds$. $\boldsymbol{v^0}$ and $\boldsymbol{u^0}$ stands respectively for the pre-strain and pre-curvature vectors.

From the previous geometrical description, we introduce the stretch and curvature vectors $\boldsymbol{v}(s)$and $\boldsymbol{u}(s)$ such that:

$$\partial_s \boldsymbol{r}(s) = \boldsymbol{v}(s) \tag{1}$$

$$\partial_s R(s) = \tilde{u}(s) R(s) \tag{2}$$

$\tilde{u}$ is the skew-symmetric matrix generated by vector $\boldsymbol{u}$. It represents the cross product in matrix notation such that $\boldsymbol{a} \times \boldsymbol{b} = \tilde{a}\boldsymbol{b}$. From any vector $\boldsymbol{a} = [a_1 \quad a_2 \quad a_3]^T$, the skew-symmetric matrix $\tilde{a}$ is defined from (3).

$$\tilde{a} = \begin{bmatrix} 0 & -a_3 & a_2 \\ a_3 & 0 & -a_1 \\ -a_2 & a_1 & 0 \end{bmatrix} \tag{3}$$

Vector $\boldsymbol{u}$ represents the curvatures along the column vectors of $R = [\boldsymbol{d_1} \quad \boldsymbol{d_2} \quad \boldsymbol{d_3}]$ where $\boldsymbol{d_1}, \boldsymbol{d_2}$ and $\boldsymbol{d_3}$are the three directors of the classical Cosserat theory which orient a material section. $\boldsymbol{v}(s)$and $\boldsymbol{u}(s)$ are the generalized strains.

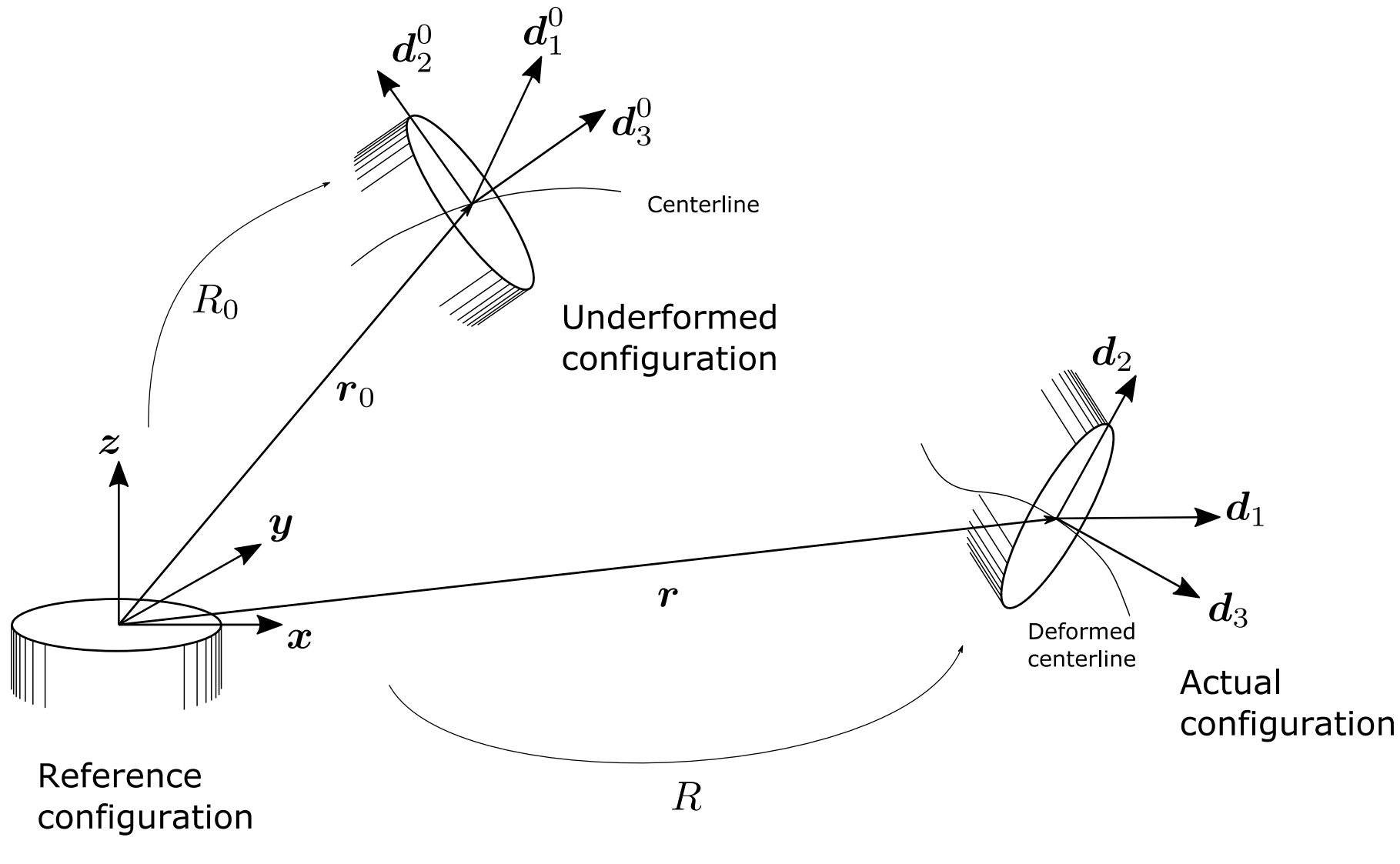

**Figure 1: Rod kinematics**

### 2.2 Balance Equations

The forces and moments acting on a generic segment $(a, s)$ where $0 < a < s < L$ are defined as $\boldsymbol{n}^+(s)$ and $\boldsymbol{m}^+(s)$ the contact (internal) force and moment exerted by segment $(s, L)$ on $(a, s)$ and $-\boldsymbol{n}^-(a)$ and $-\boldsymbol{m}^-(a)$ the contact (internal) force and moment exerted by segment $(0, a)$ on $(a, s)$. It is assumed that all other body forces and moments are of the form $\int_a^s \boldsymbol{f}(\xi)d\xi$ and $\int_a^s \boldsymbol{l}(\xi)d\xi$. $\boldsymbol{f}(s)$ and $\boldsymbol{l}(s)$ stand for the distributed forces and moments per unit length acting on the rod. The static balance equation on segment $(a, s)$ is written as follow:

$$\boldsymbol{n}^+(s) - \boldsymbol{n}^-(a) + \int_a^s \boldsymbol{f}(\xi)d\xi = 0 \tag{4}$$

$$\boldsymbol{m}^+(s) - \boldsymbol{m}^-(a) + \boldsymbol{r}(s) \times \boldsymbol{n}^+(s) - \boldsymbol{r}(a) \times \boldsymbol{n}^-(a) + \int_a^s \boldsymbol{r}(\xi) \times \boldsymbol{f}(\xi) + \boldsymbol{l}(\xi)d\xi = 0 \tag{5}$$

Holding for all $0 < a < s < L$. The required continuity (iii) imposes $\boldsymbol{n}^+(a) = \boldsymbol{n}^-(a) = \boldsymbol{n}(a)$ and $\boldsymbol{m}^+(a) = \boldsymbol{m}^-(a) = \boldsymbol{m}(a)$.

Supposing an elementary segment of length $ds$, sign convention for $\boldsymbol{n}$ and $\boldsymbol{m}$ is then: positive when acting from a material point situated at $s + ds$ on a material point at $s$.

Finally by differentiating (4) and (5) and making use of (1) the static equilibrium equation is obtained:

$$\partial_s \boldsymbol{n}(s) = -\boldsymbol{f}(s) \tag{6}$$

$$\partial_s \boldsymbol{m}(s) = -\boldsymbol{l}(s) - \boldsymbol{v}(s) \times \boldsymbol{n}(s) \tag{7}$$

## 2.3 Local Equations

Equations (1), (2), (6) and (7) constitute the four differential equations of the static problem. All fields $\boldsymbol{v}$, $\boldsymbol{u}$, $\boldsymbol{n}$ and $\boldsymbol{m}$ are expressed in a *global* Cartesian system $\{\boldsymbol{x} \quad \boldsymbol{y} \quad \boldsymbol{z}\}$ such that any field $\boldsymbol{a}$ is decomposed into $\boldsymbol{a} = a_x\boldsymbol{x} + a_y\boldsymbol{y} + a_z\boldsymbol{z}$. Equations (1), (2), (6) and (7) can be written in *local* coordinates, in the director's basis $\{\boldsymbol{d_1} \quad \boldsymbol{d_2} \quad \boldsymbol{d_3}\}$ such that if $\mathbf{a} = [\mathrm{a}_1 \quad \mathrm{a}_2 \quad \mathrm{a}_3]^T$ are the coordinates in the director's basis then $\boldsymbol{a} = a_x\boldsymbol{x} + a_y\boldsymbol{y} + a_z\boldsymbol{z} = \mathrm{a}_1\boldsymbol{d_1} + \mathrm{a}_2\boldsymbol{d_2} + \mathrm{a}_3\boldsymbol{d_3}$ and $\boldsymbol{a} = R\mathbf{a}$. In the next, global fields are written in italic while local fields are not.

Previous equations written according to the local fields $\mathbf{v}$, $\mathbf{u}$, $\mathbf{n}$ and $\mathbf{m}$ are given in (8) to (11), where dependency on curvilinear abscissa $s$ has been dropped for clarity.

$$\partial_s\boldsymbol{r} = R\mathbf{v} \tag{8}$$

$$\partial_s R = R\tilde{\mathrm{u}} \tag{9}$$

$$\partial_s\mathbf{n} = -R^T\boldsymbol{f} - \mathbf{u} \times \mathbf{n} \tag{10}$$

$$\partial_s\mathbf{m} = -R^T\boldsymbol{l} - \mathbf{v} \times \mathbf{n} - \mathbf{u} \times \mathbf{m} \tag{11}$$

In (9) $\tilde{\mathrm{u}}$ is the skew-symmetric generated by vector $\mathbf{u}$. $\mathbf{u}$ refers to *material* description (or local description) while $\boldsymbol{u}$ is the *spatial* or *inertial* description. The *material* description is adopted in this paper because the constitutive law relation is easier to write directly in the director's basis, as described below.

## 2.4 Rotation matrix parameterization

Direct numerical integration of (9) leads to accumulated errors which tend to lose the orthonormality property of rotation matrix $R$. Moreover the shooting method as described in next part consists in finding the minimum of a function (optimization viewpoint) by inverting a jacobian matrix in a Newton process. The redundant parametrization of the rotation matrix can lead to inversion issues where the jacobian rows are not independent anymore.

To get round of these problems we chose to parameterize $R$ with Modified Rodrigues Parameters (MRP). MRP parametrization allows to model a rotation whose angle is included in interval $]-2\pi, 2\pi[$. As Euler Angles, MRP constitute a minimum set of three degrees of freedom but don't suffer the well-known gimbal lock kinematic singularity of the former. Moreover the shooting method requires at some step the inversion of the rotation kinematic matrix. Unfortunately the gimbal lock is the characterization of the null determinant of the rotation kinematic matrix at some configurations. On the contrary, MRP does not suffer kinematic singularity but an orientation singularity, i.e. an orientation cannot be represented. Among all three degrees of freedom parametrization, MRP has the widest angle range without singularity. Moreover the dual set of parameters, namely Shadow Modified Rodrigues Parameters (SMRP), can be used to bypass the orientation singularity.

The easiest way to obtain the MRP $\boldsymbol{\sigma} = [\sigma_1 \quad \sigma_2 \quad \sigma_3]^T$ is from an axis and angle $(\boldsymbol{e}, \theta)$ or from a stereographic projection [10] of the quaternion representation $\boldsymbol{q} = [q_0 \quad q_1 \quad q_2 \quad q_3]^T \in H$.

$$\sigma_i = e_i \tan\frac{\theta}{4} = \frac{q_i}{q_0 + 1} \tag{12}$$

The previous MRP definition highlights an orientation singularity for $\theta = 2\pi + k4\pi,\ k \in \mathbb{Z}$, meaning that MRP can handle orientation in a range of $\theta \in ]-2\pi, 2\pi[$. If one has to handle very large rotations, this range of definition may not be sufficient, hence one can use the *shadow* set of modified Rodrigues parameters namely the SMRP at the singularities:

$$\sigma_i^S = e_i \tan\frac{\theta - 2\pi}{4} = \frac{-q_i}{-q_0 + 1} \tag{13}$$

Equation (13) is obtained by noticing that the quaternion representation of a rotation matrix is not unique. Indeed it is a two to one representation from the quaternion group $H$ to the rotation group $SO(3)$ such that quaternions $\boldsymbol{q}$ and $-\boldsymbol{q}$ actually represent the same rotation.

From (13) the SMRP are defined for $\theta \in ]0,4\pi[$ which extends the range of definition of the MRP, where forbidden angles are $\theta = k4\pi,\ k \in \mathbb{Z}$. Hence an alternate switch between MRP and SMRP using (14) and (15) allow representing any rotations in space without kinematic singularity but with an orientation discontinuity at the switching point.

$$\boldsymbol{\sigma}^S = \frac{-\boldsymbol{\sigma}}{\sigma^2} \tag{14}$$

$$\boldsymbol{\sigma} = \frac{-\boldsymbol{\sigma}^S}{\sigma^{S^2}} \tag{15}$$

Here $\sigma^2 = \sigma_1^2 + \sigma_2^2 + \sigma_3^2$ is the squared norm of the MRP vector $\boldsymbol{\sigma}$ (or its shadow representation).

From either the *standard* or *shadow* representation, the rotation matrix is obtained with (16).

$$R(\boldsymbol{\sigma}) = I - \frac{4(1-\sigma^2)}{(1+\sigma^2)^2}\tilde{\sigma} + \frac{8}{(1+\sigma^2)^2}\tilde{\sigma}^{\mathbf{2}} \tag{16}$$

The inverse rotation matrix is also computed through (16) by noticing that $R^T(\boldsymbol{\sigma}) = R(-\boldsymbol{\sigma})$.

The kinematic relation of the MRP representation (or its shadow representation) with respect to the local curvature $\mathbf{u}$ is given in (17).

$$\partial_s\boldsymbol{\sigma} = \frac{1}{4}\left((1+\sigma^2)I + 2\tilde{\sigma} + 2\tilde{\sigma}^{\mathbf{2}}\right)\mathbf{u} \tag{17}$$

This kinematic relation is always defined and does not suffer kinematic singularity as soon as the MRP or SMRP themselves are not at the singular point. In order to avoid the singular point in practice, one may note that for an orientation of $\theta = \pm\pi$ far from the forbidden angles, $\sigma^2 = {\sigma^S}^2 = 1$. Then for these specific orientations $\boldsymbol{\sigma}^S = -\boldsymbol{\sigma}$. In a practical case, when performing the integration, the switching point does not have to strictly respect $\sigma^2 = {\sigma^S}^2 = 1$. Indeed for instance, one may track the condition $\sigma^2 > 1$ or ${\sigma^S}^2 > 1$ and switching at this moment respectively to the SMRP or the MRP. Figure 2 shows a proposed switching algorithm based on [10].

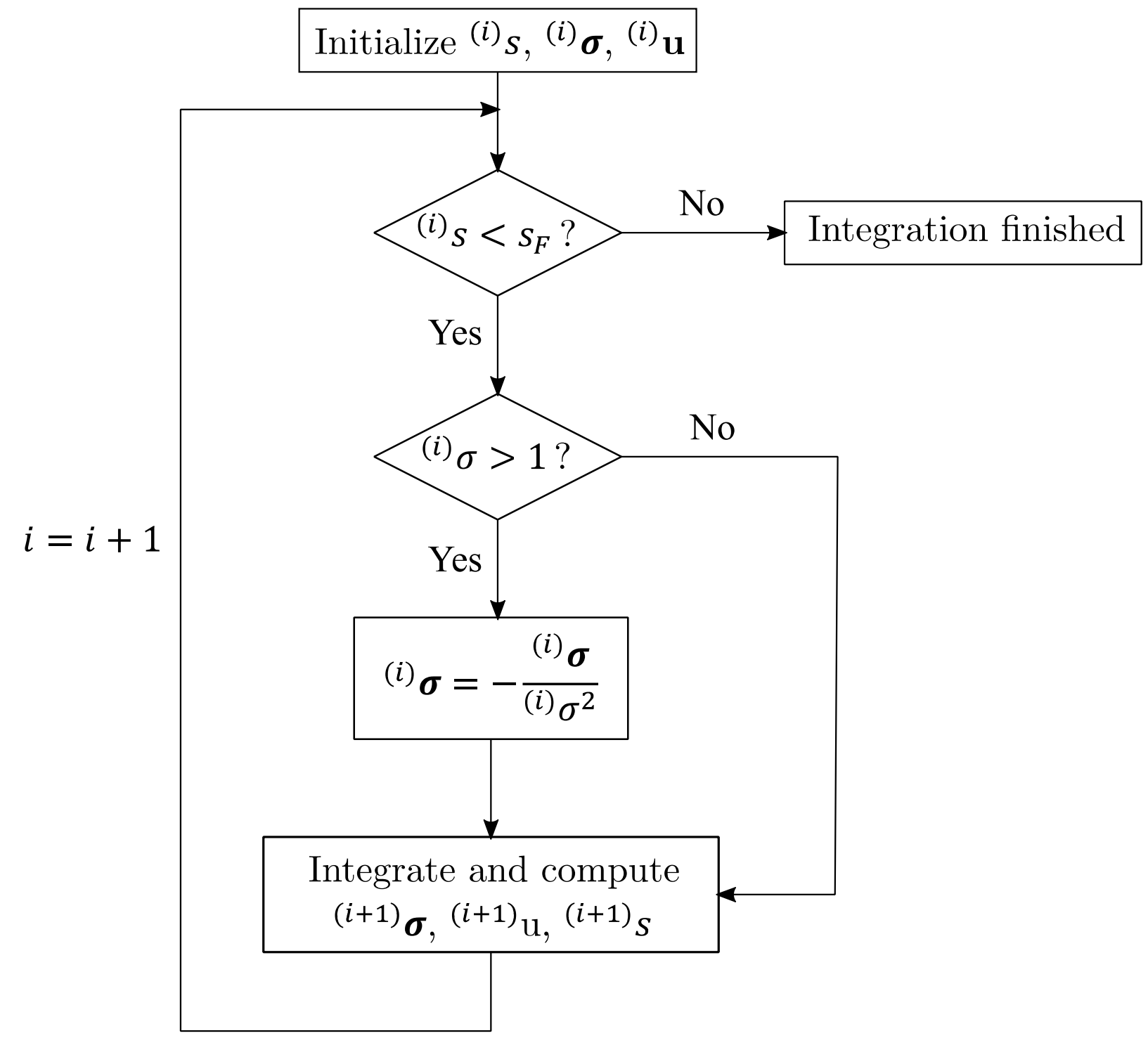

**Figure 2: MRP/SMRP switching algorithm**

From [11], the equivalent of (17), for a 1-2-3 Euler angles parameterization $\boldsymbol{a} = [\phi \quad \theta \quad \psi]^T$ such that for $R = R_{123}(\boldsymbol{a}) = R_1(\phi)R_2(\theta)R_3(\psi)$, is provided in (18).

$$\partial_s \boldsymbol{a} = \frac{1}{\cos\theta}\begin{bmatrix} \cos\phi\sin\theta & \sin\phi\sin\theta & \cos\theta \\ -\sin\phi\cos\theta & \cos\phi\cos\theta & 0 \\ \cos\phi & \sin\phi & 0 \end{bmatrix}\mathbf{u} \tag{18}$$

Kinematic matrix of (18) is singular for $\theta = \frac{\pi}{2} + k\pi$, it is the gimbal lock phenomena. Two main categories of Euler angles exist: symmetric and asymmetric. Symmetric sequences have the same first and third rotation axis, i.e. 3-1-3, 2-1-2, 1-2-1, etc… Asymmetric sequences have three different rotation axis: 1-2-3, 3-2-1, 2-1-3, etc…

All symmetric Euler angles kinematic matrix are singular when the second angle take the values of $\theta = k\pi$. On the contrary the gimbal lock for symmetric sequences occurs when the second angle is $\theta = \frac{\pi}{2} + k\pi$. Switching between two Euler angles conventions is obviously possible, but less systematic because each convention has its own kinematic relation and the switching angle condition is less convenient than for MRP and SMRP.

Finally, ODE of (9) is replaced by (17) for the rotation integration.

### *2.5 Constitutive equation*

From [9] a rod is called elastic if there are constitutive functions $\hat{\mathbf{n}}$ and $\hat{\mathbf{m}}$ such that $\mathbf{n}(\mathrm{s}) = \hat{\mathbf{n}}(\mathbf{v}(\mathrm{s}), \mathbf{u}(\mathrm{s}), s)$ and $\mathbf{m}(\mathrm{s}) = \hat{\mathbf{m}}(\mathbf{v}(\mathrm{s}), \mathbf{u}(\mathrm{s}), s)$. In the present case we will suppose a linear elasticity behavior such that (19) and (20) stand.

$$\hat{\mathbf{n}}(\mathbf{v}(s), \mathbf{u}(s), \mathrm{s}) = K^n(s)(\mathbf{v} - \mathbf{v^0}) \tag{19}$$

$$\hat{\mathbf{m}}(\mathbf{v}(s), \mathbf{u}(s), \mathrm{s}) = K^m(s)(\mathbf{u} - \mathbf{u^0}) \tag{20}$$

With the two stiffness matrices defined as

$$K^n(s) = \begin{bmatrix} G(s)A(s) & 0 & 0 \\ 0 & G(s)A(s) & 0 \\ 0 & 0 & E(s)A(s) \end{bmatrix}$$

$$K^m(s) = \begin{bmatrix} E(s)I_1(s) & 0 & 0 \\ 0 & E(s)I_2(s) & 0 \\ 0 & 0 & G(s)C(s) \end{bmatrix}$$

Where $G(s)$, $E(s)$, $A(s)$, $I_k(s)$ and $C(s)$ stand respectively for the shear modulus, the young modulus, the cross section area, the cross section moment of inertia along director $\boldsymbol{d_k}$ and the polar moment of inertia at torsional cross section's center supposed located at neutral fiber. The dependency of cross section properties with regards to curvilinear abscissa $s$ has been deliberately kept in order to emphasis that arbitrary cross sections can be modeled through the present formulation.

From (19) and (20) the strains $\mathbf{v}(s)$and $\mathbf{u}(s)$ can be recast in (21) and (22) using the inverse functions $\hat{\mathbf{v}}(\mathbf{n}, \mathbf{m}, s)$ and $\hat{\mathbf{u}}(\mathbf{n}, \mathbf{m}, s)$ and the compliance matrices $Q^n = K^{n^{-1}}$ and $Q^m = K^{m^{-1}}$.

$$\mathbf{v} = \hat{\mathbf{v}}(\mathbf{n}, \mathbf{m}, s) = \mathbf{v^0} + Q^n(s)\mathbf{n} \tag{21}$$

$$\mathbf{u} = \hat{\mathbf{u}}(\mathbf{n}, \mathbf{m}, s) = \mathbf{u^0} + Q^m(s)\mathbf{m} \tag{22}$$

Note that $\mathbf{v^0}$ and $\mathbf{u^0}$ can depend on curvilinear abscissa without loss of generality and constitute pre-requested fields. Inserting the last two equations in (10) and (11) yield (23) and (24).

$$\partial_s \mathbf{n} = -R^T \boldsymbol{f} - \hat{\mathbf{v}}(\mathbf{n}, \mathbf{m}, s) \times \mathbf{n} \tag{23}$$

$$\partial_s \mathbf{m} = -R^T \boldsymbol{l} - \hat{\mathbf{v}}(\mathbf{n}, \mathbf{m}, s) \times \mathbf{n} - \hat{\mathbf{u}}(\mathbf{n}, \mathbf{m}, s) \times \mathbf{m} \tag{24}$$

We deliberately keep functions $\hat{\mathbf{v}}(\mathbf{n}, \mathbf{m}, s)$ and $\hat{\mathbf{u}}(\mathbf{n}, \mathbf{m}, s)$ in equations (23) and (24) to highlight that the presented formalism is not restricted to linear elastic constitutive laws with constant cross section properties along the curvilinear abscissa. Indeed one has the possibility to model various elastic constitutive laws by replacing (21) and (22) by appropriate relations if a hyperelastic behavior is required for instance.

### *2.6 Boundary conditions*

In order to solve (8), (17), (23) and (24), boundary conditions (BCs) must be given. For such problems, four main boundary conditions types exist: Cauchy, Dirichlet, Neumann and Robin boundary conditions. A Cauchy BC fully defines the state vector and its derivative for one value of the state parameter (curvilinear abscissa, time, etc…) and thus define an IVPs. In the case of a TPBVP, the others three types describe BCs defined at both end of the problem. The Dirichlet BC defines the state vector, while the Neumann BC states its derivative at both ends. The Robin BC defines the state vector at one end, and its derivative at the other. The method of resolution presented here has the advantage of allowing to use any and eventually a mix of above BCs. As illustrated in the validation section, boundary conditions (a), (b), (c) and (d) of Table 1 will be considered. Others are reminded for exhaustiveness.

When internal load or moment is constrained, one can derive the BC with forces and moments depending on the kinematics. For example the case (l) derives the prismatic joint in a linear spring. All derived cases other than (l) are not reminded as they are straightforward.

**Table 1: BCs relative kinematics as unknowns and constraints**

| | Joint type | Free $\boldsymbol{X}$ | Constrained $\boldsymbol{Y}$ |
|---|---|---|---|
| (a) | Clamp | $\boldsymbol{n}$ $\boldsymbol{m}$ | $\boldsymbol{r}$ $\boldsymbol{\sigma}$ |
| (b) | Spherical (sphere/sphere) | $\boldsymbol{\sigma}$ $\boldsymbol{n}$ | $\boldsymbol{r}$ $\boldsymbol{m}$ |
| (c) | Imposed Force | $\boldsymbol{r}$ $\boldsymbol{\sigma}$ | $\boldsymbol{n}$ $\boldsymbol{m}$ |
| (d) | Sphere/Cylinder of axis $\boldsymbol{x}$ | $x$ $\boldsymbol{\sigma}$ $n_y$ $n_z$ | $y$ $z$ $n_x$ $\boldsymbol{m}$ |
| (e) | Prismatic of axis $\boldsymbol{x}$ | $x$ $n_y$ $n_z$ $\boldsymbol{m}$ | $y$ $z$ $\boldsymbol{\sigma}$ $n_x$ |
| (f) | Punctual of axis $\boldsymbol{z}$ | $x$ $y$ $n_z$ $\boldsymbol{\sigma}$ | $z$ $n_x$ $n_y$ $\boldsymbol{m}$ |
| (g) | Cylinder of axis $\boldsymbol{x}$/Plane axis $\boldsymbol{z}$ | $x$ $y$ $\sigma_x$ $\sigma_z$ $n_z$ $m_y$ | $z$ $\sigma_y$ $n_x$ $n_y$ $m_x$ $m_z$ |
| (h) | Pivot of axis $\boldsymbol{x}$ | $\sigma_x$ $\boldsymbol{n}$ $m_y$ $m_z$ | $\boldsymbol{r}$ $\sigma_y$ $\sigma_z$ $m_x$ |
| (i) | Sliding pivot of axis $\boldsymbol{x}$ | $x$ $\sigma_x$ $n_y$ $n_z$ $m_y$ $m_z$ | $y$ $z$ $\sigma_y$ $\sigma_z$ $n_x$ $m_x$ |
| (j) | Plane/Plane of axis $\boldsymbol{z}$ | $x$ $y$ $\sigma_z$ $n_z$ $m_x$ $m_y$ | $z$ $\sigma_x$ $\sigma_y$ $n_x$ $n_y$ $m_z$ |
| (k) | Screw joint of axis $\boldsymbol{x}$ | $x$ $n_y$ $n_z$ $m_x$ $m_y$ $m_z$ | $y$ $z$ $\sigma_x(x)$ $\sigma_y$ $\sigma_z$ $n_x(m_x)$ |
| (l) | Linear spring of axis $\boldsymbol{x}$ | $x$ $n_y$ $n_z$ $\boldsymbol{m}$ | $n_x(x)$ $y$ $z$ $\boldsymbol{\sigma}$ |
| ⋮ | ⋮ | ⋮ | ⋮ |

In a multibody approach, a kinematical joint is a representation of a physical linkage which imposes and/or frees relative motions between two connected bodies. A key point in such description is the duality between the kinematic and the static torsors of such linkages. In this work we consider perfect frictionless mechanical linkage as physical joints to impose the boundary conditions. That is to say the power of the linkage internal contact forces must be null. The power of the contact forces is computed through the comoment of the kinematic and the static torsor. Then when a kinematic degree of freedom is constrained, the corresponding relative velocity is null, but the contact force is undetermined. In this case we will speak about free static degree of freedom. On the contrary, when a motion is kinematically free, it is statically constrained with null corresponding force or moment components. For example, in the case of a spherical joint, no relative translational velocities between the connected bodies are allowed. Therefore the relative position vector is constant in space and no moment can be transmitted at the reduction point from the kinematic joint. However rotations are free to evolve, as well as internal forces.

The TPBVP of static rods is written with regards to global coordinates $\boldsymbol{r}(s)$ and $\boldsymbol{\sigma}(s)$ of the material section. Therefore, integral form of admissible relative velocities constraints are used to impose the fields $\boldsymbol{r}(s)$ and $\boldsymbol{\sigma}(s)$ at the boundaries.

Let $\boldsymbol{X}$ and $\boldsymbol{Y}$ respectively designate the free and constrained fields related to either the relative positions, orientations, forces or moments of a kinematical joint as defined in Table 1. Boundary conditions constrain both position and rotation vectors as well as internal forces and moments. Then one can evaluate vector $\boldsymbol{\varphi}(s) = [\boldsymbol{r}(s), \boldsymbol{\sigma}(s), \mathbf{n}(s), \mathbf{m}(s)]^T$ at boundaries $s = 0$ and $s = L$ such that (25) holds.

$$\boldsymbol{\varphi} = \boldsymbol{\gamma}(\boldsymbol{X}, \boldsymbol{Y}) \tag{25}$$

Function $\boldsymbol{\gamma}$ defines a map from the relative boundary condition admissible kinematics to the set of state vector. Hence the boundary state vectors $\boldsymbol{\varphi}_{s_0}$ and $\boldsymbol{\varphi}_{s_L}$ can be computed from free kinematic $\boldsymbol{X}_{s_0}$, $\boldsymbol{X}_{s_L}$ and constrained $\boldsymbol{Y}_{s_0}$, $\boldsymbol{Y}_{s_L}$ variables such that $\boldsymbol{\varphi}_{s_0,s_L} = \boldsymbol{\gamma}_{s_0,s_L}(\boldsymbol{X}_{s_0,s_L}, \boldsymbol{Y}_{s_0,s_L})$. Functions $\boldsymbol{\gamma}_{s_0}$ and $\boldsymbol{\gamma}_{s_L}$ are respectively the mappings for the joints situated at $s = 0$ and $s = L$.

One can also define the inverse maps from the state vectors of the TPBVP of static rods such that $\boldsymbol{\Gamma}_{\boldsymbol{X}}$ and $\boldsymbol{\Gamma}_{\boldsymbol{Y}}$ extract the unknown and constrained fields of the joint corresponding to assemblies of relative positions, orientation, forces and moments.

$$\boldsymbol{X} = \boldsymbol{\Gamma}_{\boldsymbol{X}}(\boldsymbol{\varphi}) \tag{26}$$

$$\boldsymbol{Y} = \boldsymbol{\Gamma}_{\boldsymbol{Y}}(\boldsymbol{\varphi}) \tag{27}$$

In (26) and (27), mapping $\boldsymbol{\Gamma}_{\boldsymbol{X}}$ and $\boldsymbol{\Gamma}_{\boldsymbol{Y}}$ are valued functions defined on the set of vectors state parameter $\boldsymbol{\varphi}$. Notations from previous paragraph can be reflected such that if $\boldsymbol{\Gamma}_{\boldsymbol{X}_{s_0}}$, $\boldsymbol{\Gamma}_{\boldsymbol{Y}_{s_0}}$ and $\boldsymbol{\Gamma}_{\boldsymbol{X}_{s_L}}$, $\boldsymbol{\Gamma}_{\boldsymbol{Y}_{s_L}}$ are the mappings for the joints situated at $s = 0$ and $s = L$, then $\boldsymbol{X}_{s_0,s_L} = \boldsymbol{\Gamma}_{\boldsymbol{X}_{s_0,s_L}}(\boldsymbol{\varphi}_{s_0,s_L})$ and $\boldsymbol{Y}_{s_0,s_L} = \boldsymbol{\Gamma}_{\boldsymbol{Y}_{s_0,s_L}}(\boldsymbol{\varphi}_{s_0,s_L})$.

Finally the pre-mentioned ODEs are recast in a first order ODE (28), with the boundary conditions constraint equations (29) and (30). The boundary conditions formalism as presented here describes indifferently position/rotation, imposed forces/moment or a mix of boundary conditions. Note that in the TPBVP the boundary conditions are always separated.

$$\partial_s \boldsymbol{\varphi}(s) = \boldsymbol{F}(\boldsymbol{\varphi}, s) \tag{28}$$

$$\boldsymbol{C}_{s_0}\left(\boldsymbol{\varphi}(s=0), \overline{\boldsymbol{Y}}_{s_0}\right) = \mathbf{0} \tag{29}$$

$$\boldsymbol{C}_{s_L}\left(\boldsymbol{\varphi}(s=L), \overline{\boldsymbol{Y}}_{s_L}\right) = \mathbf{0} \tag{30}$$

Notation $\bar{\cdot}$ is used here to distinguish a constrained field variable $\boldsymbol{Y}$ from its known imposed values $\overline{\boldsymbol{Y}}$ resulting from the joint type. $F(\boldsymbol{\varphi}, s)$ is computed from (8), (17), (23) and (24) and summarized in (31).

$$\boldsymbol{F}(\boldsymbol{\varphi}, s) = \begin{cases} R^T\left(\mathbf{v}^{\mathbf{0}} + Q^n \mathbf{n}\right) \\ \frac{1}{4}\left((1+\sigma^2)I + 2\tilde{\sigma} + 2\tilde{\sigma}^2\right)\left(\mathbf{u}^{\mathbf{0}} + Q^m \mathbf{m}\right) \\ -R^T \boldsymbol{f} - \left(\mathbf{u}^{\mathbf{0}} + Q^m \mathbf{m}\right) \times \mathbf{n} \\ -R^T \boldsymbol{l} - \left(\mathbf{v}^{\mathbf{0}} + Q^n \mathbf{n}\right) \times \mathbf{n} - \left(\mathbf{u}^{\mathbf{0}} + Q^m \mathbf{m}\right) \times \mathbf{m} \end{cases} \tag{31}$$

Basically, the static rod problem being a TPBVP, (28) to (30) is solved using $\boldsymbol{\varphi}_{s_0}$ and $\boldsymbol{\varphi}_{s_L}$ evaluated through (25) and Table 1 depending on the physical configuration considered.

Equations (29) and (30) describe the most general separated boundary conditions constraint equations definition. In most cases, constraints that imposes full vectors fields $\mathbf{n}$, $\mathbf{m}$ or $\boldsymbol{r}$ are expressed from (32) where $\boldsymbol{\Gamma}_{\boldsymbol{Y}_{s_0,s_L}}$ is defined in (27). This is the case for clamp, spherical and imposed force joints.

$$\boldsymbol{C}_{s_0,s_L}\big(\boldsymbol{\varphi}(s=0,L),\boldsymbol{\varphi}_{s_0,s_L}\big)=\boldsymbol{\Gamma}_{\boldsymbol{Y}_{s_0,s_L}}\big(\boldsymbol{\varphi}(s=0,L)\big)-\overline{\boldsymbol{Y}}_{s_0,s_L} \tag{32}$$

In the case of mixed boundary conditions mixing components of positions, orientation, forces and moments, special care must be taken to the basis in which components are written. Mechanical linkages often define an attached basis defining preferred axis for the relative translations or rotations motions. In these cases, constraints functions $\boldsymbol{C}_{s_0,s_L}$ can be nonlinear composite functions. For instance in case of a punctual joint of axis $\boldsymbol{z}$, the constraint imposing zero $n_x$ relative force would be enforced by $\boldsymbol{C}_{s_0,s_L}=\Big(R\big(\boldsymbol{\sigma}(s=0,L)\big)\,\mathbf{n}(s=0,L)\Big)\cdot\boldsymbol{x}=n_x=0$.

The same remark holds when imposing full rotations. Indeed, one may not impose directly $\boldsymbol{\sigma}(s=0,L)$ but a composed rotation with the imposed rotation. If $R\big(\boldsymbol{\sigma}_{s_0,s_L}\big)$ is the imposed rotation matrix at tips, the composed rotation $\boldsymbol{\sigma}'$ from the composed rotation of (33) is constrained to zero. This case holds when all rotations are constrained, i.e. for clamp, prismatic, etc... linkages.

$$R(\boldsymbol{\sigma}')=R^T\big(\boldsymbol{\sigma}_{s_0,s_L}\big)R\big(\boldsymbol{\sigma}(s)\big) \tag{33}$$

$R(\boldsymbol{\sigma}')$ can be indifferently computed through rotation matrix multiplication using the MRP to rotation matrix relation of (16) or through direct MRP rotation composition of (34) considering the composition $R(\boldsymbol{\sigma}')=R(\boldsymbol{\sigma}'')R(\boldsymbol{\sigma})$.

$$\boldsymbol{\sigma}'=\frac{\big(1-\sigma''^2\big)\boldsymbol{\sigma}+(1-\sigma^2)\boldsymbol{\sigma}''+2\boldsymbol{\sigma}\times\boldsymbol{\sigma}''}{1+\sigma''^2\sigma^2-2\boldsymbol{\sigma}''\cdot\boldsymbol{\sigma}} \tag{34}$$

Equation (34) is not defined when the final rotation $\boldsymbol{\sigma}'$ match the forbidden angle of the orientation singularity. To avoid that, depending on the highest norm, switching $\boldsymbol{\sigma}$ or $\boldsymbol{\sigma}''$ to their SMRP representation is sufficient to guarantee the well MRP multiplication and $\boldsymbol{\sigma}'$ to be in the unit sphere, i.e. $\sigma'^2 < 1$.

Processing of partially imposed rotations such as pivot is more subtle. Let vectors $\boldsymbol{d_i}(\boldsymbol{\sigma}) = R(\boldsymbol{\sigma}).\boldsymbol{e}_i$ with $\boldsymbol{e}_i = \{\boldsymbol{x}, \boldsymbol{y}, \boldsymbol{z}\}$ be the local orientation of the joint. Suppose the rotation free to rotate around the prescribed unit vector axis $\boldsymbol{d}_k$. Unknown fields are then the angle of rotation $\theta_{\boldsymbol{d}_k}$ around axis $\boldsymbol{d}_k$, the two moment components orthogonal to the plane defined by $\boldsymbol{d}_k$ and the full internal force vector $\boldsymbol{n}$. The angle can be retrieved from the MRP parameterization $\boldsymbol{\sigma}' = \boldsymbol{d}_k \tan\frac{\theta_{\boldsymbol{d}_k}}{4}$ and the two moment components are $m_{\boldsymbol{d}_{j\neq k}} = \boldsymbol{m}.\boldsymbol{d}_{j\neq k}$ such that $\boldsymbol{m}_{\perp \boldsymbol{a}} = m_{\boldsymbol{d}_{j\neq k}}\boldsymbol{d}_{j\neq k} = \boldsymbol{m} - (\boldsymbol{m}.\boldsymbol{a}).\boldsymbol{a} = -\tilde{a}^2\boldsymbol{m}$. $\boldsymbol{\sigma}'$ is here the MRP representation of the rotation composition previously introduced in (33). The constraints in the pivot case are the imposed moment value along the rotation axis $\boldsymbol{d}_k.\boldsymbol{m} = \overline{m}_{0,L}$, the two constraints on the axis of rotation: $\boldsymbol{\sigma}'.\boldsymbol{d}_{j\neq k} = 0$ and the three positions $\boldsymbol{x} = \overline{\boldsymbol{x}}_{0,L}$.

## 3. SHOOTING METHOD

### *3.1 Single shooting method*

The shooting method consists in transforming the TPBVP of (28), (29) and (30) into the search of the roots of a nonlinear constraint function being evaluated with a succession of IVPs.

The main point of the classical single shooting method is to impose $\boldsymbol{Y}_{s_0}$ to its known values $\overline{\boldsymbol{Y}}_{s_0}$ and *guess* the unknown part $\boldsymbol{X}_{s_0}$ in order to build an initial state vector $\boldsymbol{\varphi}_{s_0} = \boldsymbol{\gamma}_{s_0}\left(\boldsymbol{X}_{s_0}, \boldsymbol{Y}_{s_0} = \overline{\boldsymbol{Y}}_{s_0}(\boldsymbol{X}_{s_0})\right) = \boldsymbol{\gamma}_{s_0}(\boldsymbol{X}_{s_0})$ from (25). Here we didn't assume $\overline{\boldsymbol{Y}}_{s_0}$ fully independent with regards to $\boldsymbol{X}_{s_0}$.

Indeed in some situation $\overline{\boldsymbol{Y}}_{s_0}$ explicitly depends on $\boldsymbol{X}_{s_0}$. For instance a screw joint would have a dependency between the angle rotated and the translation through the pitch of the joint as shown in case (k) of Table 1.

Integrating this initial state $\boldsymbol{\varphi}_{s_0}$ through the ODE allows to evaluate the final state $\boldsymbol{\varphi}(s = L; \boldsymbol{\varphi}_{s_0}) = \boldsymbol{\varphi}\left(L; \boldsymbol{\gamma}_{s_0}(\boldsymbol{X}_{s_0})\right)$ which depends on the initial state $\boldsymbol{\varphi}_{s_0} = \boldsymbol{\gamma}_{s_0}(\boldsymbol{X}_{s_0})$. This final state is compared to final boundary condition imposed values through the inverse mapping of (27) such that by defining $\boldsymbol{Y}_{s_L}(\boldsymbol{X}_{s_0}) = \boldsymbol{\Gamma}_{\boldsymbol{Y}_{s_L}}\left(\boldsymbol{\varphi}\left(L; \boldsymbol{\gamma}_{s_0}(\boldsymbol{X}_{s_0})\right)\right)$, the constraints function $\boldsymbol{C}\left(\boldsymbol{Y}_{s_L}(\boldsymbol{X}_{s_0}), \overline{\boldsymbol{Y}}_{s_L}\left(\boldsymbol{X}_{s_L}(\boldsymbol{X}_{s_0})\right)\right)$ of (35) is nullified. In the same way as before, $\overline{\boldsymbol{Y}}_{s_L}$ may depend on the current values of unknown part of the boundary condition evaluated at $s = L$, namely $\boldsymbol{X}_{s_L}(\boldsymbol{X}_{s_0}) = \boldsymbol{\Gamma}_{\boldsymbol{X}_{s_L}}\left(\boldsymbol{\varphi}\left(L; \boldsymbol{\gamma}_{s_0}(\boldsymbol{X}_{s_0})\right)\right)$.

In this description only $\boldsymbol{X}_{s_0}$ is assumed unknown at the starting joint and constraint function $\boldsymbol{C}$ depends only on $\boldsymbol{X}_{s_0}$.

One can extend the former formalism by supposing both $\boldsymbol{X}_{s_0}$and $\boldsymbol{Y}_{s_0}$ unknown. This leads to the same beginning steps but $\boldsymbol{Y}_{s_0}$ is also guessed. The ODE is integrated in the same way. However the constraint function is augmented and replaced by (36). The constraint function $\boldsymbol{C}$ now depends on both $\boldsymbol{X}_{s_0}$ and $\boldsymbol{Y}_{s_0}$.

$$\begin{aligned} \boldsymbol{C}(\boldsymbol{X}_{s_0}) &= \boldsymbol{C}\left(\boldsymbol{Y}_{s_L}(\boldsymbol{X}_{s_0}), \overline{\boldsymbol{Y}}_{s_L}\left(\boldsymbol{X}_{s_L}(\boldsymbol{X}_{s_0})\right)\right) \\ &= \boldsymbol{C}_{s_L}\left(\boldsymbol{\varphi}(L; \boldsymbol{\varphi}_{s_0}), \overline{\boldsymbol{Y}}_{s_L}(\boldsymbol{X}_{s_0})\right) \end{aligned} \tag{35}$$

$$\begin{aligned} \boldsymbol{C}(\boldsymbol{X}_{s_0}, \boldsymbol{Y}_{s_0}) &= \boldsymbol{C}\left(\boldsymbol{X}_{s_0}, \boldsymbol{Y}_{s_0}, \overline{\boldsymbol{Y}}_{s_0}(\boldsymbol{X}_{s_0}), \boldsymbol{Y}_{s_L}(\boldsymbol{X}_{s_0}), \overline{\boldsymbol{Y}}_{s_L}\left(\overline{\boldsymbol{X}}_{s_L}(\boldsymbol{X}_{s_0})\right)\right) \\ &= \begin{cases} \boldsymbol{C}_{s_0}\left(\boldsymbol{\varphi}(s = 0; \boldsymbol{\varphi}_{s_0}), \overline{\boldsymbol{Y}}_{s_0}(\boldsymbol{X}_{s_0})\right) \\ \boldsymbol{C}_{s_L}\left(\boldsymbol{\varphi}(s = L; \boldsymbol{\varphi}_{s_0}), \overline{\boldsymbol{Y}}_{s_L}(\boldsymbol{X}_{s_0})\right) \end{cases} \end{aligned} \tag{36}$$

Whatever the method employed between assuming the initial integrated vector partially pre-constrained or not, the unknowns of constraint function are generalized in a single variable vector noted $\boldsymbol{Z}$. In the former case $\boldsymbol{Z} = \boldsymbol{X}_{s_0}$, in the latter $\boldsymbol{Z} = \left[\boldsymbol{X}_{s_0}, \boldsymbol{Y}_{s_0}\right]^{\boldsymbol{T}}$. From (26) and (27) we also define the function $\boldsymbol{Z} = \boldsymbol{\Gamma}_{\boldsymbol{Z}}(\boldsymbol{\varphi})$ without ambiguity.

The shooting problem in its general form is written as follow:

$$RF: \begin{cases} IVP: \begin{cases} \partial_s \boldsymbol{\varphi}(s) = \boldsymbol{F}(\boldsymbol{\varphi}, s) \\ \boldsymbol{\varphi}(s = 0) = \boldsymbol{\varphi}_{s_0} = \boldsymbol{\gamma}(\boldsymbol{Z}) \end{cases} \\ \boldsymbol{C}(\boldsymbol{Z}) = \boldsymbol{0} \end{cases} \quad (37)$$

$\boldsymbol{C}$ is a constraint vector function. $\boldsymbol{\varphi}_{s_0}$ can be fully or partially unknown and depends on variables $\boldsymbol{Z}$ which constitute the unknowns of the Root Finding (RF) algorithm. The shooting method principle is then to iterate over the guessed part $\boldsymbol{Z}$ of the IVP in order to fulfill the constraint function $\boldsymbol{C}$ evaluated through successive integrations of the ODE to evaluate $\boldsymbol{Y}_{s_L}$ (and $\boldsymbol{Y}_{s_0}$ if necessary). The resolution process is described in Figure 3.

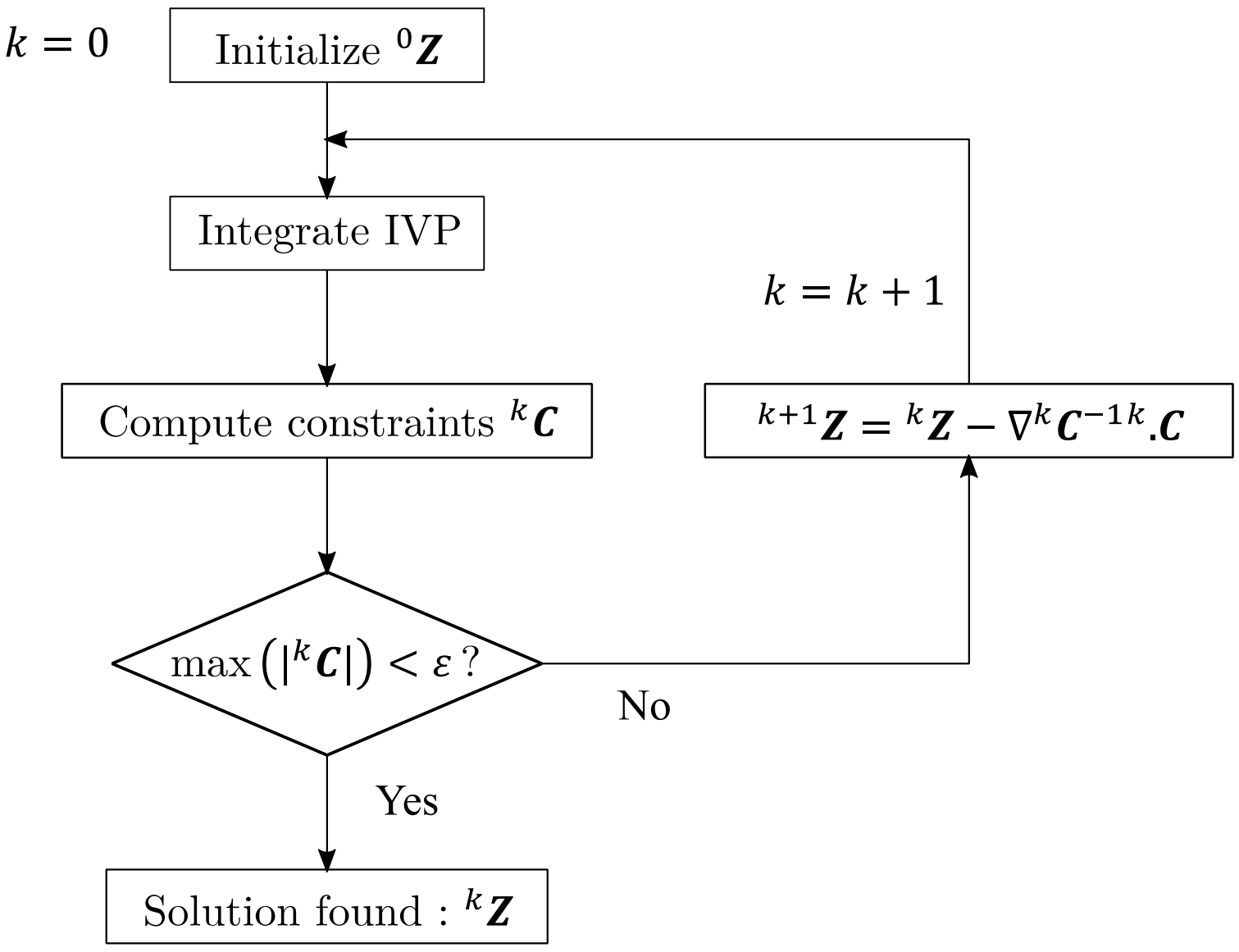

**Figure 3: Shooting method resolution process**

In this paper, the continuous integration of the ODE will be dealt with an embedded Runge-Kutta-Fehlberg 4-5 (RKF45) scheme, together with the rotation switching algorithm described in Figure 2. The roots of the constraint equation are sought by means of a Newton-Raphson algorithm, where the gradient is computed through finite differences. In order to solve the problem with Newton algorithm, the size of $\boldsymbol{C}$ is restricted by the condition $\dim \boldsymbol{C} = \dim \boldsymbol{Z}_{s_0}$ and $\nabla\boldsymbol{C}$ is supposed fully ranked $\dim \ker \nabla\boldsymbol{C} = 0$. This condition is necessary for the Jacobian to be invertible in the Newton process. In the general case, this last condition is not necessarily obtained. Indeed in some occasions the configuration is determined to within a rigid motion. In order to avoid the rank deficient case and invert the Jacobian properly, a SVD decomposition and inversion is performed.

Concerning the use of an adaptive integrator, the RKF45 absolute error parameter corresponds to the cumulated error between the two schemes. It follows that adapting integration steps all over the segment, leads to a cumulated absolute error between a RK4 and a RK5 scheme never exceeding prescribed integration absolute tolerance. This let us be confident about the precision of integration at segment's end. In addition the number of integration points is kept to a minimum while reaching the required accuracy.

In a general point of view, the shooting method as described inherits the drawbacks of nonlinear root seeking algorithms: it may not converge if the first guess of ${}^{0}\boldsymbol{Z}$ is too far from the sought root of $\boldsymbol{C}$. This notion of "distance" strongly relies on the physical equations involved, the $\boldsymbol{C}$ shape and the conditioning of $\nabla\boldsymbol{C}$. Stiff problems - meaning that small perturbations of $\boldsymbol{Z}$ propagated through the IVP induce large variations of $\boldsymbol{C}$ - restrain the convergence radius of the root finding algorithm, necessitating ${}^{0}\boldsymbol{Z}$ to be closer to the solution. Moreover the final solution depends on the first initialization guessed for the Newton algorithm and represents one solution in the set of all possible zeros of the nonlinear vector constraint function $\boldsymbol{C}$. In order to avoid convergence problems, the multiple shooting method may be used [2].

Unlike discretized approaches, a great strength of the present shooting method is to keep continuous all processed fields and parameters. For instance if the linear elastic constitutive law of (21) and (22) are used, the compliance matrices can arbitrarily depend on the state parameter $s$. Hence this formalism allows for instance the modeling of a smooth continuous change of shape of the cross section.

For the case of rods and in order to reduce the unknowns number of the Newton root finding, we choose to constrain the part of $\boldsymbol{\varphi}_{s_0}$ such that we fall in the first case $\boldsymbol{Z} = \boldsymbol{Z}_{s_0} = \boldsymbol{X}_{s_0}$ described above. Indeed, because we parametrized the rotation with a minimal representation, half of the unknowns vector $\boldsymbol{\varphi}(s) = [\boldsymbol{r}(s), \boldsymbol{\sigma}(s), \mathbf{n}(s), \mathbf{m}(s)]^T$ at $s = 0$, say $\boldsymbol{\varphi}_{s_0}$, is known. Thus, we can choose to use the other half part (i.e. only unconstrained fields of the BC instead of all fields) as Newton's unknowns in order to keep the dimensions of Jacobian $\nabla\boldsymbol{C}$ as small as possible, enabling a quicker inversion in the root finding process. Hence, within this formulation ${}^{0}\boldsymbol{Z}$ and $\boldsymbol{C}$ have the same size of only 6. The single shooting problem is given in (38).

$$\begin{cases} \begin{cases} \partial_s \boldsymbol{\varphi}(s) = \boldsymbol{F}(\boldsymbol{\varphi}, s) \\ \boldsymbol{\varphi}_{s_0} = \boldsymbol{\gamma}_{s_0}(\boldsymbol{Z}_{s_0}) \end{cases} \\ \boldsymbol{C}(\boldsymbol{Z}_{s_0}) = \boldsymbol{0} \end{cases} \tag{38}$$

Moreover, in the present description, the unknown part $\boldsymbol{Z}_{s_0}$ of $\varphi_{s_0}$ depends on the boundary condition chosen at $s = 0$. Table 1 summarizes the known and unknown parts of each joint type. It is highlighted that the Jacobian of this problem is only a 6x6 matrix although all geometrical nonlinearities are taken into account. Inverting a 6x6 matrix is then very effective compared to nonlinear discretized approaches where the Jacobian size depends on the squared number of elements.

## *3.2 Multiple shooting method*

The multi-shooting method [12] is based on the exact same system as (37) and differs from the single shooting method on the size of $\boldsymbol{\varphi}$, $\boldsymbol{Z}$ and $\boldsymbol{C}$.

The first application purpose of the multi-shooting method is when the IVP has a bad conditioning [2]. Indeed, the principle is to cut the initial segment into $n$ sub-segments following the same differential equations, adding new sets of unknowns and new sets of constraints obtained through continuity conditions. This leads to smaller integrated intervals, which tend to improve the conditioning of $\nabla \boldsymbol{C}$ at the cost of increasing the size of the problem to solve. Integration and constraints expressed on smaller sub intervals can be viewed as a linearization technique of the constraint function $\boldsymbol{C}$ here. Indeed smaller integration ranges induce less variations of the propagated state vector $\boldsymbol{\varphi}_{s_L}$ hence of the constraint function $\boldsymbol{C}$.

In our case, we propose another application purpose of the multi-shooting technique which is the handling of discontinuities in fields $\boldsymbol{r}(s)$, $\boldsymbol{\sigma}(s)$, $\mathbf{n}(s)$ and $\mathbf{m}(s)$ because one assumption of the governing equations was to impose sufficiently regular fields according to hypothesis (iii).

Discontinuities can be: a punctual external force at given curvilinear abscissa, a change of property on the material section (different sections geometries, different constitutive laws) or different sets of distributed forces applied along the segments. A set of validation cases is based on the use of the multiple shooting method to handle discontinuities with regard to external distributed and punctual forces.

Whatever the application purpose is, the principle of the multi-shooting for solving TPBVP is based on the same idea of solving several initial value problems on smaller subdomains than the initial domain of definition of the one dimensional parameter of the problem (i.e. the curvilinear abscissa). While dividing into subintervals, new unknowns are generated as well as new constraints to match the interval wise trajectories on the edges of the curvilinear abscissa subdomains.

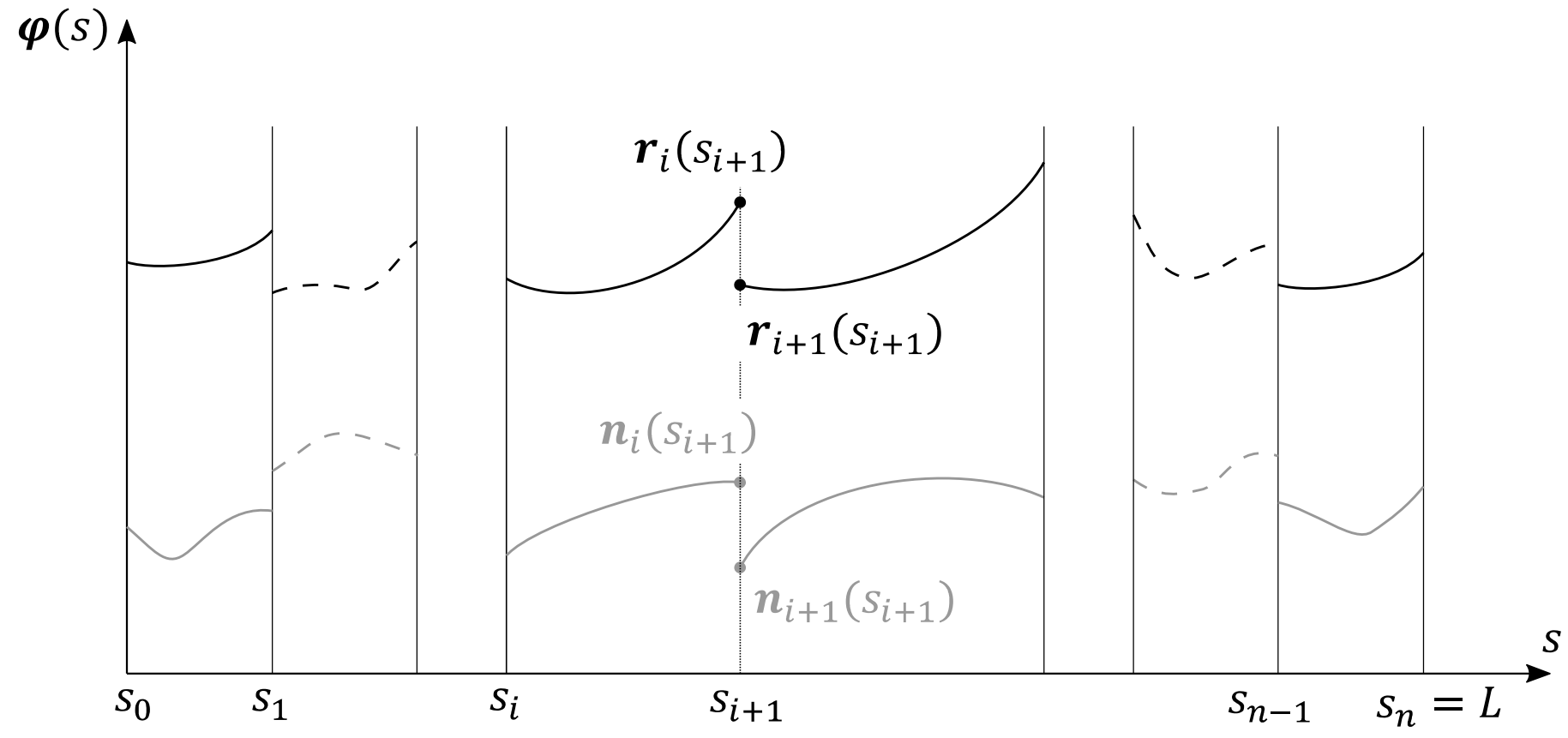

**Figure 4: Illustration of the multiple shooting method**

Figure 4 illustrates such subdivision for a chosen curvilinear abscissa decomposition $s_0 = 0 < s_1 < s_i < s_{i+1} < s_n = s_L = L$. The solution on the $i\ th$ interval is noted $\boldsymbol{\varphi}_i(s)$. At the beginning of the $i + 1\ th$ subinterval, the full vector is unknown and constitutes additional variables (39).

$$\boldsymbol{\varphi}_{i+1}(s = s_{i+1}) = \boldsymbol{\gamma}_{s_{i+1}}\left(\boldsymbol{Z}_{s_{i+1}}\right) \tag{39}$$

Moreover at the end of $i\ th$ subinterval, the integrated values must be equal to the values at the beginning of the $i+1\ th$. Hence the constraint vector $\boldsymbol{C}$ is augmented with the new constraints $\boldsymbol{C}_{s_i}\left(\boldsymbol{\Gamma}_{\boldsymbol{Z}_{s_{i+1}}}(\boldsymbol{Z}_i), \boldsymbol{Z}_{s_{i+1}}\right) = \boldsymbol{0}$ where $\boldsymbol{\Gamma}_{\boldsymbol{Z}_{s_{i+1}}}(\boldsymbol{Z}_{s_i}) = \boldsymbol{\Gamma}_{\boldsymbol{Z}_{s_{i+1}}}\left(\boldsymbol{\varphi}_i\left(s = s_{i+1}; \boldsymbol{Z}_{s_i}\right)\right)$ is the $i\ th$ subinterval integrated vector fully constrained at final curvilinear abscissa $s_{i+1}$. Thus in the case of a multiple shooting method functions $\boldsymbol{\gamma}_{s_i}$ and $\boldsymbol{\Gamma}_{\boldsymbol{Z}_{s_i}}$ are identity functions such that $\boldsymbol{\gamma}_{s_{i+1}}\left(\boldsymbol{Z}_{s_{i+1}}\right) = \boldsymbol{Z}_{s_{i+1}}\ \forall i \in ]0, n[$. $\boldsymbol{C}_{s_i}$ is expressed such that the imposed constraint is equivalent to a clamp joint from (32) and (33).

Finally by introducing $\boldsymbol{\phi}(s) = [\boldsymbol{\varphi}_0, \cdots, \boldsymbol{\varphi}_i, \cdots, \boldsymbol{\varphi}_{n-1}]^T$; (38) is recast in (42) which as the same structure as (37) as already mentioned.

$$\boldsymbol{G}(\boldsymbol{\phi}, s) = \begin{bmatrix} \boldsymbol{F}(\boldsymbol{\varphi}_0, s) \\ \vdots \\ \boldsymbol{F}(\boldsymbol{\varphi}_i, s) \\ \vdots \\ \boldsymbol{F}(\boldsymbol{\varphi}_{n-1}, s) \end{bmatrix} \tag{40}$$

$$\boldsymbol{H}\left(\boldsymbol{Z}_{s_0}, \cdots, \boldsymbol{Z}_{s_i}, \cdots, \boldsymbol{Z}_{s_{n-1}}\right) = \begin{bmatrix} \boldsymbol{C}_{s_1}\left(\boldsymbol{\Gamma}_{\boldsymbol{Z}_{s_1}}\left(\boldsymbol{Z}_{s_0}\right), \boldsymbol{Z}_{s_1}\right) \\ \vdots \\ \boldsymbol{C}_{s_i}\left(\boldsymbol{\Gamma}_{\boldsymbol{Z}_{s_i}}\left(\boldsymbol{Z}_{s_{i-1}}\right), \boldsymbol{Z}_{s_i}\right) \\ \vdots \\ \boldsymbol{C}_{s_L}\left(\boldsymbol{Y}_{s_L}\left(\boldsymbol{Z}_{s_{n-1}}\right), \overline{\boldsymbol{Y}}_{s_L}\left(\boldsymbol{Z}_{s_{n-1}}\right)\right) \end{bmatrix} \tag{41}$$

$$\begin{cases} \begin{cases} \partial_s \boldsymbol{\phi}(s) = \boldsymbol{G}(\boldsymbol{\phi}, s) \\ \boldsymbol{\phi}_0 = \left[\boldsymbol{\gamma}_{s_0}\left(\boldsymbol{Z}_{s_0}\right) \quad \cdots \quad \boldsymbol{\gamma}_{s_i}\left(\boldsymbol{Z}_{s_i}\right) \quad \cdots \quad \boldsymbol{\gamma}_{s_{n-1}}\left(\boldsymbol{Z}_{s_{n-1}}\right)\right]^T \end{cases} \\ \boldsymbol{H}\left(\boldsymbol{Z}_{s_0}, \cdots, \boldsymbol{Z}_{s_i}, \cdots, \boldsymbol{Z}_{s_{n-1}}\right) = \boldsymbol{0} \end{cases} \tag{42}$$

The IVP system of (42) can be integrated in a raw or by subdomain. In the first case, one may normalize the curvilinear abscissa such that each sub-segment lies in a common integration domain. Such variable change can be done by introducing the translated curvilinear abscissa $s^i$ of the $i\ th$ interval obtained from the global curvilinear abscissa $s$ with (43). The global curvilinear abscissa referring to the original segment is given by the inverse relation in (44).

$$s^i = \frac{s - s_i}{s_{i+1} - s_i} \in [0,1] \tag{43}$$

$$s = s_i + (s_{i+1} - s_i)s^i \tag{44}$$

Equation (44) let us recompute the curvilinear abscissa referring to the original segment. This is necessary to evaluate the cross section constitutive functions and the associated constitution laws of (21) and (22), which are most of the time defined from the curvilinear abscissa origin of the uncut segment.

In (42) we also made the choice to construct $\boldsymbol{\phi}_0$ with the known part of the boundary condition $\overline{\boldsymbol{Y}}_{s_0}$extracted from Table 1 such that $\boldsymbol{Z}_{s_0} = \boldsymbol{X}_{s_0}$. As described in the single shooting method paragraph, we could have made the choice to use $\boldsymbol{Z}_{s_0} = \left[\boldsymbol{X}_{s_0}, \boldsymbol{Y}_{s_0}\right]^{\boldsymbol{T}}$ instead and appending the constraint $\boldsymbol{C}_{s_0}$ to $\boldsymbol{H}$, without loss of generality on the multiple shooting resolution.

## *3.3 Multiple shooting method as multibody approach*

The present work now develops the multiple shooting method from a multibody point of view. This special seeing is motivated by the principle argument that in the classical multiple shooting method formalism, it is possible, but not generic to model more than two segments connecting at the same point. For instance in the three segments configuration proposed in [7], special constraint equations have been added to deal with the middle connected node. We develop in this part a particular model based on the multibody philosophy of static bodies linked together with kinematic connections, but where segments are solved using the shooting method.

Figure 5 presents three configurations: (a) a classical multibody open chain, (b) the multiple shooting method as presented in the last section and (c) the proposed mixed multibody/multiple shooting approach.

In configuration (a), body $j$ is connected to body $j-1$ through any standard kinematic joint $k-1$ and to body $j+1$ through any other kinematic joint $k$. $\boldsymbol{r}_j$ and $R_j$ respectively describe absolute position and rotation degrees of freedom of body $j$. $\boldsymbol{r}_{j//j+1}$ and $R_{j//j+1}$ respectively describe relative position and rotation of body $j$ with respect to body $j+1$.

The first idea here is to consider a slender element as a particular body that can be linked through kinematic joints as shown in diagram (b) of Figure 5. The multiple shooting method described in the previous paragraph is a specific case of this modeling in which only clamp linkages are considered. As previously said, the limitation of this modeling is that a segment can be connected to only one other segment.

Finally if slender elements are considered as bodies, they can be linked to rigid bodies as shown in diagram (c) of Figure 5. This particular approach encompasses the lumped mass modeling, where slender elements are actually discretized in rigid bodies linked together with linear springs.

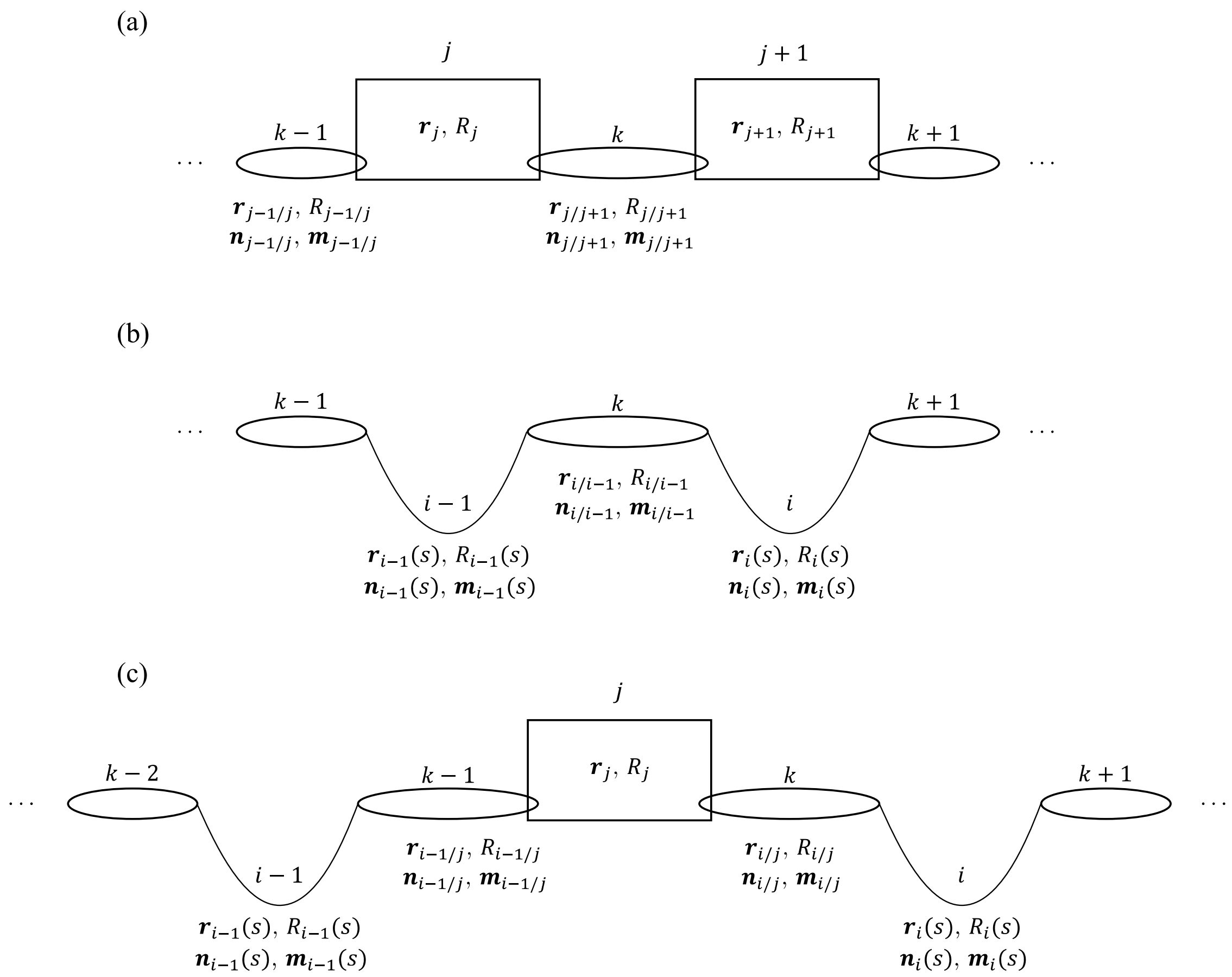

**Figure 5: Multibody and multiple shooting kinematic chains**

In the formalisms (b) and (c), joints constitute boundary conditions for slender elements and considerations of the previous boundary conditions paragraph still apply. Moreover the formalism of (b) has been developed in the previous multiple shooting method paragraph.

The structure of (c) let model arbitrary line assemblies by means of linking the rigid body $j$ with any required joints. As in (a), variables $\boldsymbol{r}_{i/j+1}$, $R_{i/j+1}$, $\boldsymbol{n}_{i/j+1}$ and $\boldsymbol{m}_{i/j+1}$ constitute the internal unknowns of joint $k$. These variables are to be brought closer to the one described in the boundary conditions description. Indeed they are the entry point of the root finding stage of the single shooting method. Moreover the positions $\boldsymbol{r}_j$ and orientation $R_j$ of each rigid body are additional unknowns of the root seeking algorithm as mentioned in the previous paragraph. The additional constraints come from two sources: constraints from joints as described in above paragraphs and from (45). Indeed (45) represents the equilibrium of all forces and moments acting on the rigid body $j$ with the relations $\boldsymbol{n}_{j+1/j} = -\boldsymbol{n}_{j/j+1}$ and $\boldsymbol{m}_{j+1/j} = -\boldsymbol{m}_{j/j+1}$. Moments from different connections $\boldsymbol{m}_{p/j}$ must obviously be expressed at the same chosen reduction point.

$$\left.\begin{array}{l}\sum_p \boldsymbol{n}_{p/j} \\ \sum_p \boldsymbol{m}_{p/j}\end{array}\right\} = 0 \tag{45}$$

In the presented (c) formalism, the rigid bodies positions and orientations variables are considered as new unknowns and balance equation of (45) constitutes new constraints to be appended to the global constraint vector $\boldsymbol{C}$.

In the previously mentioned multiple shooting method, each sub-segment arrive at and start from the rigid body through a clamp joint. The arrival connection $k-1$ imposes the segment's end and the rigid body positions and orientation to be equal such that $\boldsymbol{r}_{i-1} - \boldsymbol{r}_j = \boldsymbol{r}_{i-1/j} = 0$ and $R_{i-1}R_j^T = R_{i-1/j} = I$. For rotations the MRP convention can be adopted, the constraint equation being $\boldsymbol{\sigma}_{i-1/j} = \boldsymbol{0}$. The starting connection $k$ generates the root seeking algorithm unknowns $\boldsymbol{n}_{i/j+1}$ and $\boldsymbol{m}_{i/j+1}$. As in formalism (b) of previous section, in the (c) case cutting a segment in two sub-segments generates 6 root seeking unknowns as well as 6 new constraints. It is reminded that the joint generating constraints is arbitrary and is only determined by the choice of curvilinear abscissa origin. Nevertheless whatever the origin chosen for segment integration, unknowns and constraints types generated from joints may be different but their number will remain the same. For the multiple shooting case, the equation system to solve is given in (42).

Finally the presented (c) formalism extends the capability of the classical multiple shooting method by allowing the modeling of assemblies of several segments connected to the same rigid body as well as other connection types than clamp equivalent constraints.

## 4. NUMERICAL EXAMPLES

The described single shooting and multiple shooting methods are confronted to several cases studied in the literature in order to demonstrate the validity and the accuracy of the proposed approach. These tests have several objectives. The first one is to test the ability of the shooting and multiple shooting method to solve efficiently nonlinear static rod problems in which several deformation modes are involved together. Analytical results will be used as soon as possible for comparison. When no analytical formulations are available, results are compared to the available results in the literature.

The second objective is to test the ability of the modified Rodrigues parameters (MRP) and its shadow set (SMRP) to handle problems involving large 3D rotations. The switching procedure is confronted to the helical example in which the orientation singularity is reached several times.

The third objective is to demonstrate the capability of the present formulation to deal with various boundary conditions.

The fourth objective is to validate the multiple shooting approach described previously.

Literature based cases 4.1 to 4.5 focus on the first two objectives. The last two objectives are covered by cases 4.6 to 4.9 based on material strength examples.

The first set of cases are well-known problems of large displacement of elastic beam structures based on a pure bending of an Euler beam with and without pre-bend as well as with and without out of plane force. Cases four and five are run in order to show the good accuracy of the solving procedure when strong couplings between shear, bending, torsion and extension are involved.

Finally the second set of validation cases gives a validation of the multi-shooting approach seen in the multi-body point of view when rods geometrical discontinuities are modeled through shooting approach.

Pre-strained shapes are modeled by the prescription of strains $\mathbf{v^0}$ and curvatures $\mathbf{u^0}$ of the unloaded configuration. Unless specifically specified, a tolerance of 1.e-8 for the Newton algorithm and an absolute RK45 error of 1.e-12 are used for all validation cases.

### *4.1 Cantilever under pure free end moment*

This example is a pure bending moment applied at the free end of an initially straight beam [13,14,15,16,17,18].

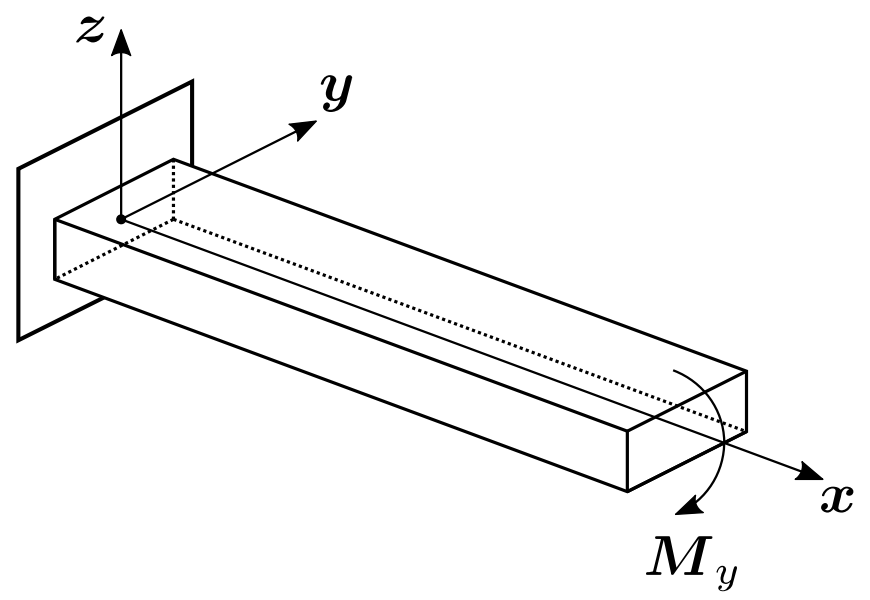

**Figure 6: Straight cantilever under pure tip moment**

The beam is directed along $\boldsymbol{x}$ axis, clamped at the origin at first end with a pure moment of magnitude $M$ is applied around $\boldsymbol{y}$ axis at the other extremity. For the configuration the solution [15] is:

$$
\begin{aligned}
m_y(s) &= M \\
x(s) &= \frac{1}{\rho} sin(\rho s) \\
z(s) &= \frac{1}{\rho} (cos(\rho s) - 1) \\
\rho &= \frac{M}{EI}
\end{aligned}
\tag{46}
$$

The following properties of the beam are taken into account:

**Table 2: Mechanical properties**

| | |
|---|---|
| Length at rest $L$ | 1 m |
| Second moment of inertia $I_1 = I_2 = I$ | 2 $m^4$ |
| Polar moment of Inertia $C$ | 1 $m^4$ |
| Cross section area $A$ | 0.25 $m^2$ |
| Young Modulus $E$ | 1 $N/m^2$ |
| Poisson's ratio $\nu$ | 0.3 |

$\rho$ is the constant curvature of the exact arc of circle solution. Numerical results are obtained using a single shooting method. The moment is parameterized by the ratio $\frac{ML}{EI}$. Five imposed moments are considered such that $\frac{ML}{EI}$ takes the values $0, \frac{\pi}{2}, \pi, \frac{3}{2}\pi$ and $2\pi$. Beam profiles obtained from numerical simulation are shown in Figure 7. Maximum absolute errors along the beam are calculated with respect to the analytical solution of (46) and summarized in Table 3. Regarding the results, absolute errors for all fields are kept really low, which demonstrates that the proposed MRP parameterization associated to the shooting method can solve beams with large deflections.

**Table 3: Straight cantilever under pure moment – Maximum absolute errors**

| | Fields | | |
|---|---|---|---|
| $\frac{ML}{EI}$ | $x$ | $z$ | $m_y$ |
| 0 | 0.00E+00 | 3.33E-16 | 0.00E+00 |
| $\frac{\pi}{2}$ | 4.58E-12 | 8.28E-12 | 2.07E-13 |
| $\pi$ | 1.10E-11 | 7.89E-12 | 4.14E-13 |
| $\frac{3}{2}\pi$ | 1.68E-11 | 6.91E-12 | 6.20E-13 |
| $2\pi$ | 1.39E-11 | 8.09E-12 | 4.08E-11 |

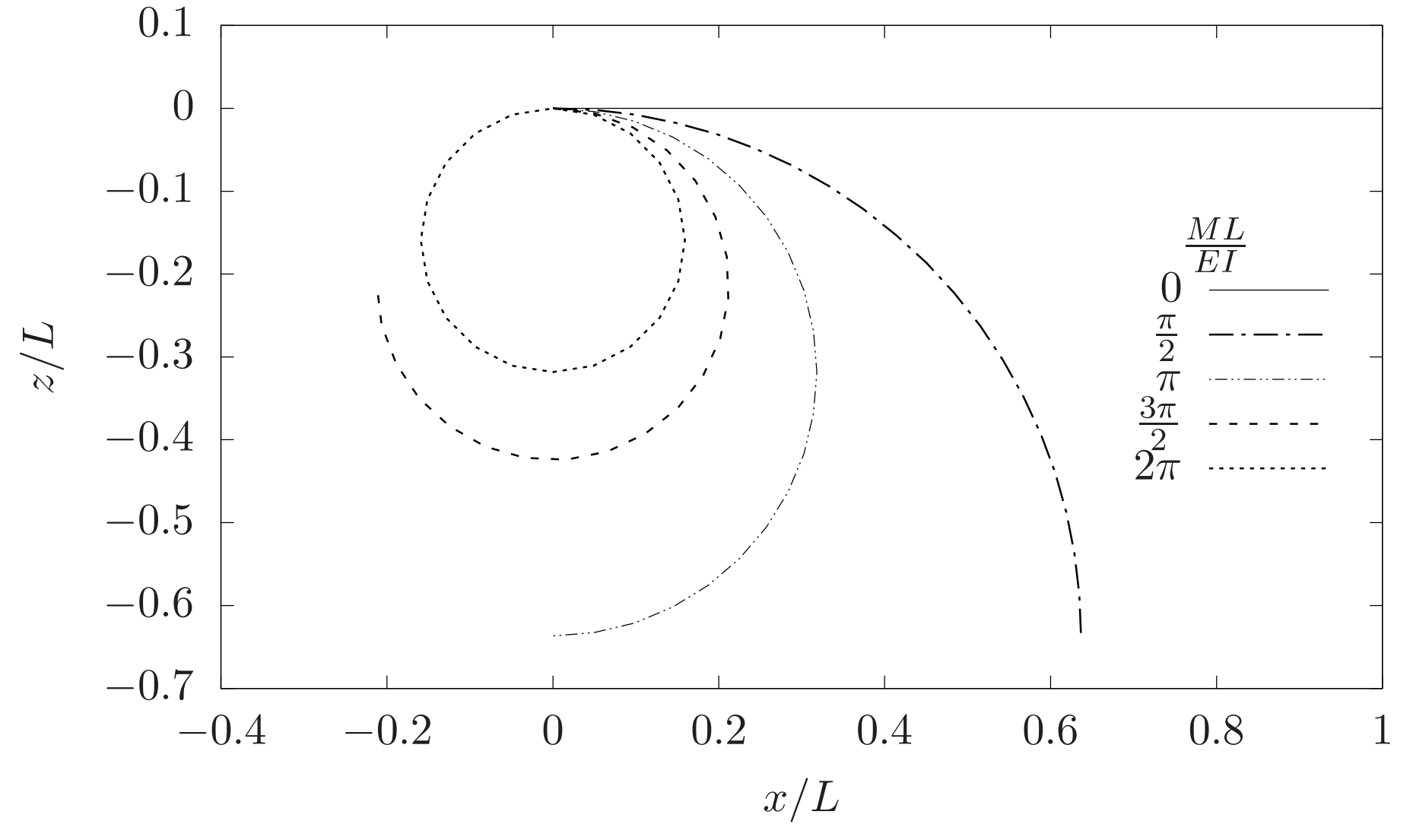

**Figure 7: Straight cantilever deflection under pure moment**

### *4.2 Pre-bent cantilever under pure free end moment*

This second case considers a pre-curved beam into a full circle on which an end moment is applied to unroll the beam [13,19]. Unrolling a circular beam let us demonstrate that the shooting method properly considers initially curved centroid axis.

This application is derived from the previous, at the difference that the beam is pre-bent with constant curvature $\rho_0$ around $\boldsymbol{z}$ axis. The analytical solution is modified such that the curvature of the beam is modified to:

$$\rho = \frac{M}{EI} + \rho_0 \tag{47}$$

The initial curvature $\rho_0 = \frac{2\pi}{L}$ is taken to be such that reference configuration (internal force and moment free configuration) of the beam is a full circle. As for the above case, the moment is parameterized by the ratio $\frac{ML}{EI}$. The physical properties given in Table 2 are also used.

Profiles obtained through numerical simulations for values of $0, \frac{\pi}{2}, \pi, \frac{3}{2}\pi$ and $2\pi$ are shown in Figure 8. Table 4 summarizes the maximum absolute errors along the beam. As for previous case, absolute errors are very low. Initial curvature are naturally handled through (21)-(22) of the shooting method equation set, while finite elements methods need specific reformulation of equilibrium equations in order to improve accuracy if the reference configuration is not straight [19,20].

**Table 4: Pre-bent cantilever under pure moment – Maximum absolute errors**

| | Fields | | |
|---|---|---|---|
| $\frac{ML}{EI}$ | $x$ | $z$ | $m_y$ |
| 0 | 1.39E-11 | 8.03E-12 | 0.00E+00 |
| $\frac{\pi}{2}$ | 1.68E-11 | 7.02E-12 | 2.07E-13 |
| $\pi$ | 1.09E-11 | 7.93E-12 | 4.14E-13 |
| $\frac{3}{2}\pi$ | 4.58E-12 | 8.35E-12 | 6.20E-13 |
| $2\pi$ | 0.00E+00 | 1.30E-12 | 4.08E-11 |

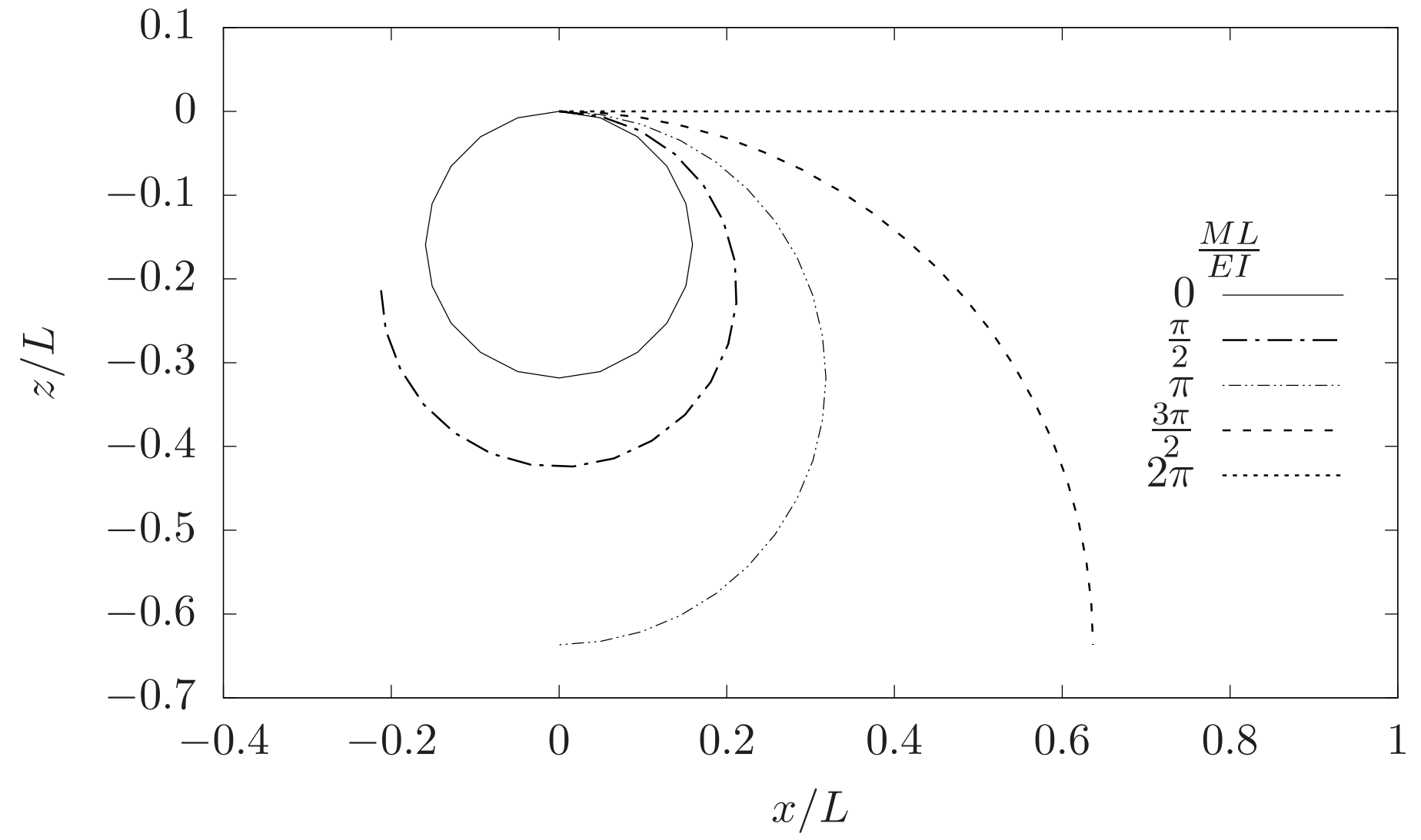

**Figure 8: Pre-bent cantilever deflection under pure moment**

### *4.3 Helical form bending*

The third case bends an initially straight beam in to a helical form. At the difference with previous cases, the cantilever beam is submitted simultaneously at its free end to a point moment and an out-of-plane point force. This case has been first studied by [17] and later by several authors [13,14,21,22]. The aim of this example is to validate the rotation parameterization with MRP. As underlined by [17], [21] as well as [22] the choice of the rotation parameterization is a crucial point in order to obtain correct results.

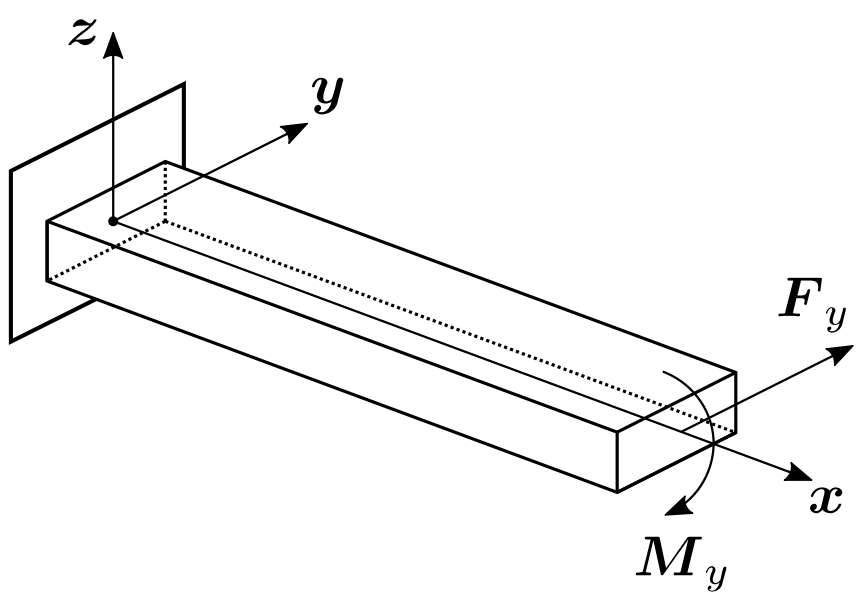

**Figure 9: Straight cantilever under tip moment and tip load**

The bending and shear stiffness are respectively $EI_1 = EI_2 = GC = 10^2$ and $EA = 2GA = 10^4$, the Poisson's ratio is taken null $\nu = 0$ and the length is $L = 10$.

The force $F = 50\lambda$ and moment $M = 200\pi\lambda$ loads are incrementally increased from $\lambda = 0$ to $\lambda = 1$. Increasing both point force and moment lead the initially straight beam to deform in a helical shape and the displacement along the force axis starts oscillating. The shooting method process and the unknowns of the Newton algorithm are initialized from the former step.

Results are shown in Figure 10 and Figure 11 where dashed line corresponds to $G = \frac{E}{2}$, solid line corresponds to $G = E$ and dot line corresponds to no shear strain (infinite shear stiffness).

Figure 13 shows results extracted from several references. Before performing any comparison, it is important to note that some references used different shear stiffness. Indeed [13,14,17,21] took $EA = GA = 10^4$ whereas [22] used $EA = 2GA = 10^4$. [23] did not specify any numerical values but it is assumed that authors took same assumption as [17], saying $EA = GA = 10^4$.

It is seen that the out-of-plane displacement are in very good accordance with [14,17]. We may notice that a slight slope on the load-displacement curve is observed when performing numerical simulations with a shear modulus being the Young modulus. If the shear modulus is taken to be half the Young modulus, as expected, the slope is steeper because the strains are higher due to the lower shear stiffness and (19).This is what is observed in [22].

It is interesting to see on Figure 10 and Figure 11 that an Euler Bernoulli beam (i.e. no shear strain or infinite shear stiffness) corresponds to a centered oscillating out-of-plane displacement around the zero value. Regarding Figure 13 it seems that the results from [13,21,23] are closer to the infinite shear solution hence the elements used in these references may induce higher stiffness despite the same shear modulus values chosen.

It is shown that shear modeling is an important feature on this specific case and has significant impact compared to a pure bending modeling.

Regarding the present results, we can conclude that the MRP parametrization is well adapted for the modeling of Cosserat rod section when large rotation kinematics coupled to shear are involved.

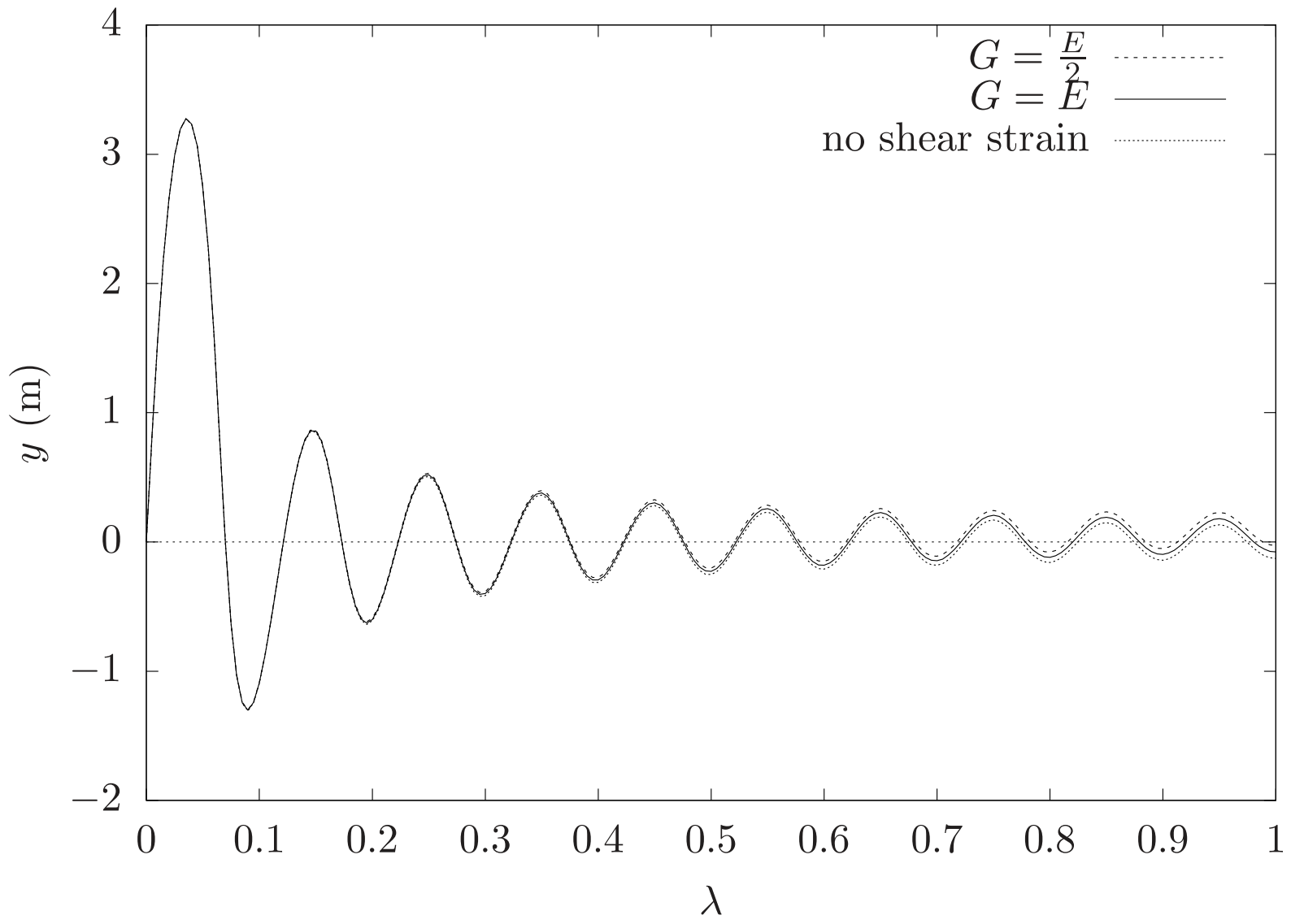

**Figure 10: Load-displacement curve for helical beam**

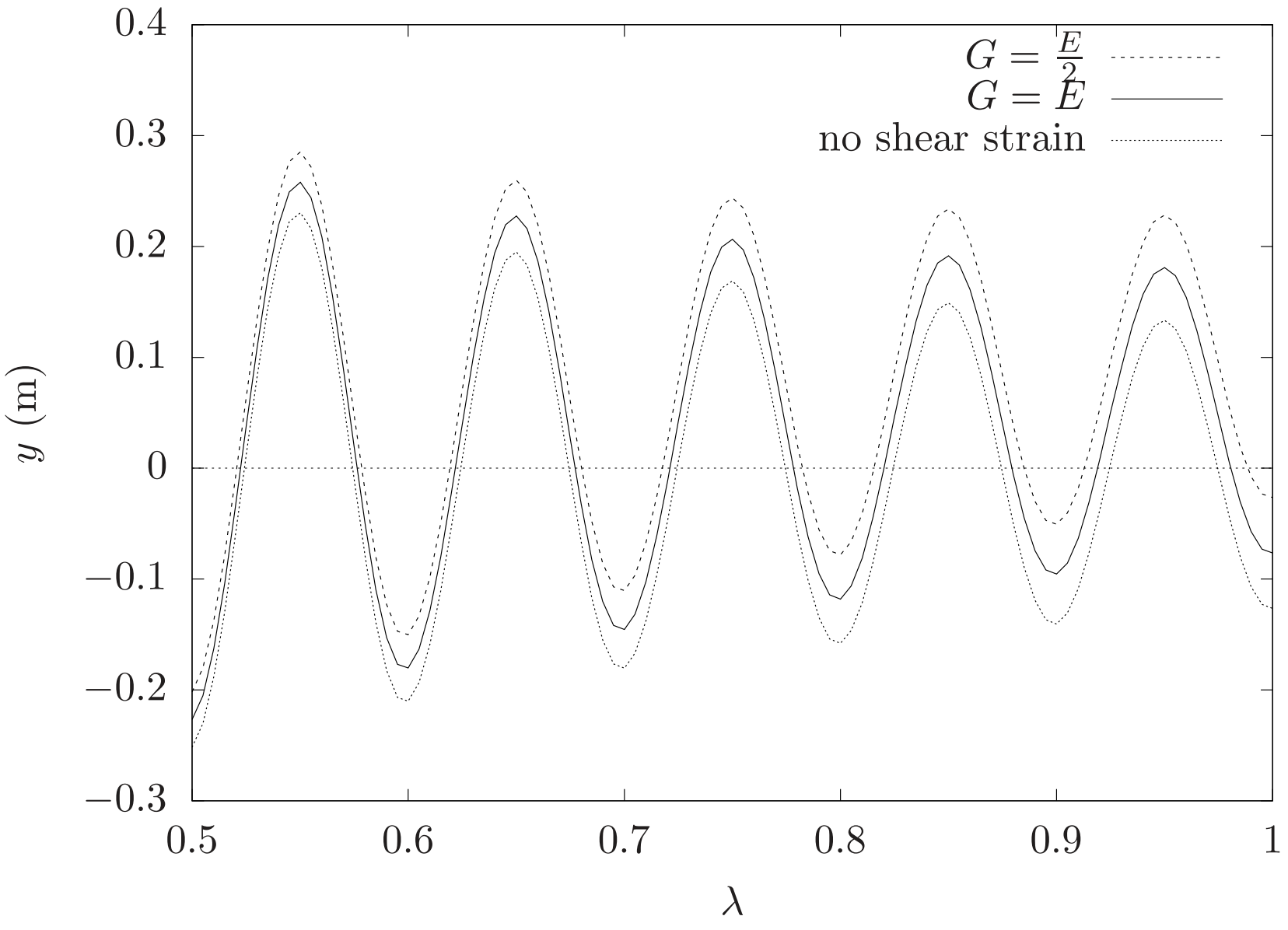

**Figure 11: Load-displacement curve for helical beam zoomed**

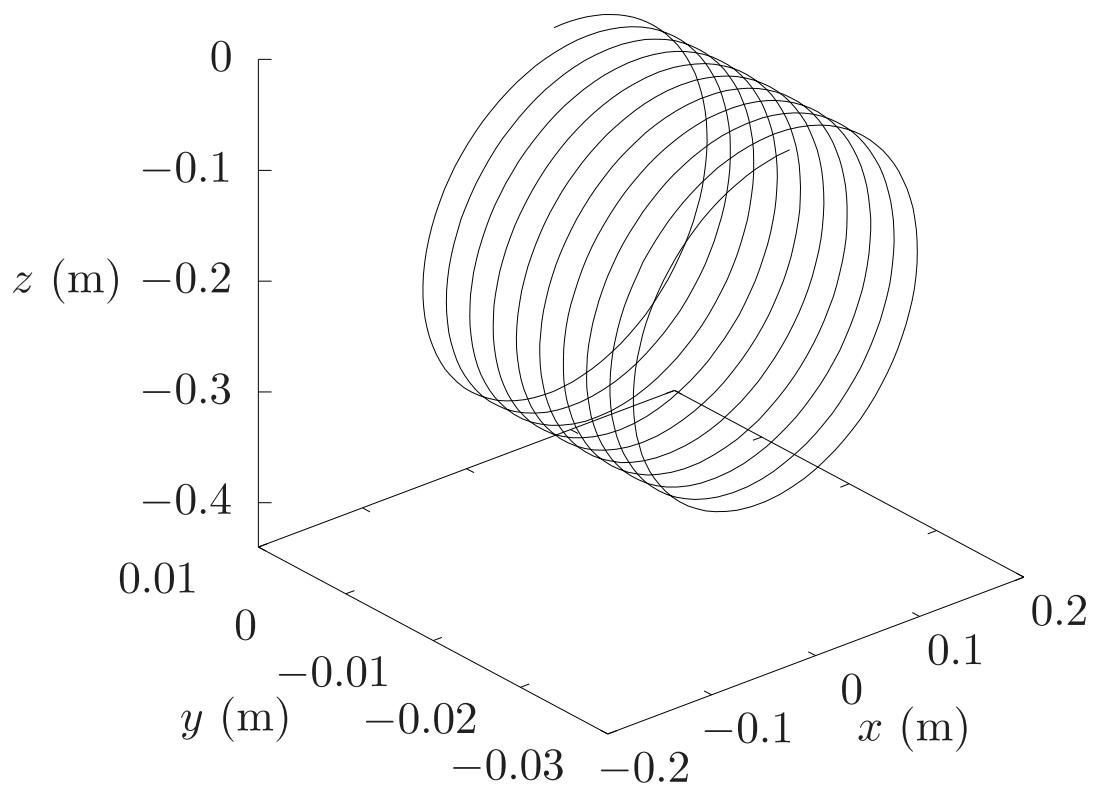

**Figure 12: Deformed shape of helical beam**

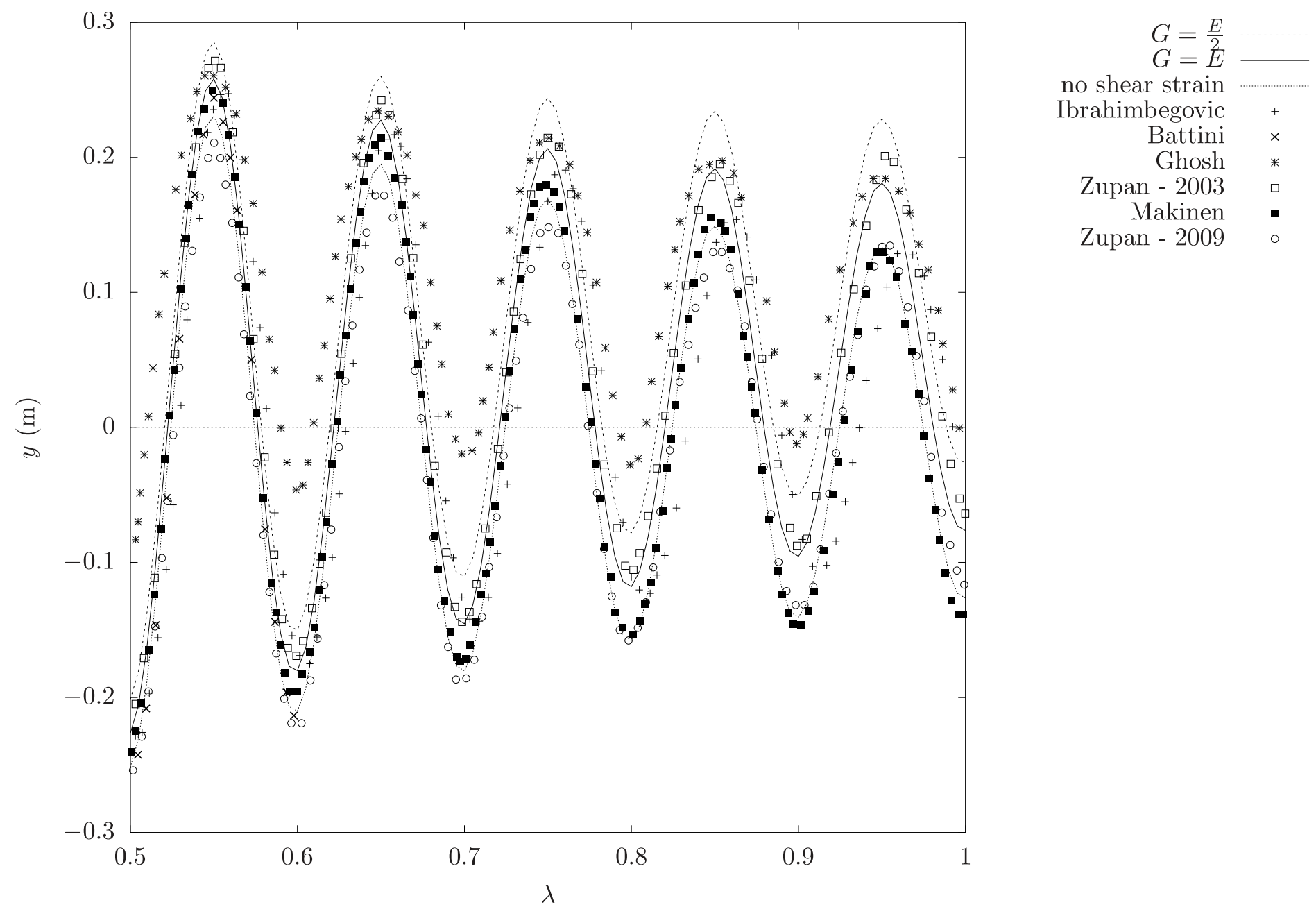

**Figure 13: Load-displacement curve for helical beam zoomed – comparison with [17] Ibrahimbegovic, [21] Battini, [22] Ghosh, [14] Zupan (2003), [23] Makinen and [13] Zupan (2009)**

## *4.4 Twisted cantilever*

The fourth case described in Figure 14 is based on an initially twisted beam first proposed by [24] reused by [13,14,25,26]. Two loads cases are considered at the free end: a unit force is applied in the thickness direction and a unit force applied in the width direction. The mechanical properties of the beam are described in Table 5. Theoretical solution derived from plate model are considered from [24] in the literature. A comparison between the results obtained with the shooting method and the literature is provided in Table 6. Results obtained with the shooting method are quantitatively identical from [13] and their quaternion-based three-dimensional finite element beam theory. Very good agreement is also found with the theoretical solution of [24]. This demonstrates the capability of the shooting method to handle accurately combinations of twist, shear and bending deformation modes. Two Newton's iterations were needed for the shooting method to converge.

**Table 5: pre-twisted beam - mechanical properties**

| | |
|---|---|
| Width $t$ | 0.32 m |
| Height $h$ | 1.1 m |
| Length | 12 m |
| Young modulus $E$ | 29.e6 N/m$^2$ |
| Poisson's ration $\nu$ | 0.22 |

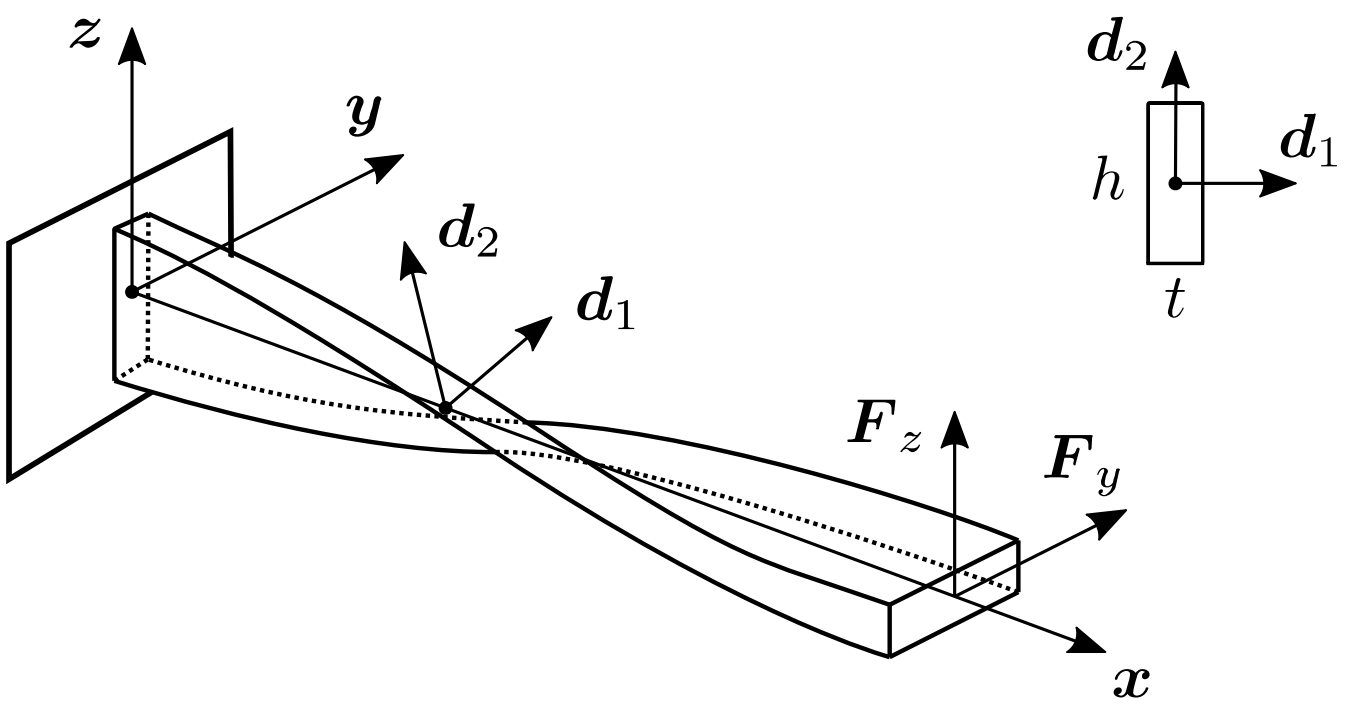

**Figure 14: pre-twisted cantilever**

**Table 6: pre-twisted cantilever submitted to bending forces**

| | Tip displacement | | | |
|---|---|---|---|---|
| | $F_y = 1N,\ F_z = 0N$ | | $F_y = 0N,\ F_z = 1N$ | |
| Reference | $y$ | $z$ | $y$ | $z$ |
| [24] MacNeal and Harer (1985) | 0.00542400 | - | - | 0.00175400 |
| [25] Dutta and White (1992) | 0.005402 | - | - | 0.001741 |
| [26] Ibrahim and Frey (1994) | 0.005411 | - | - | 0.001751 |
| [20] Zupan and Saje (Analytical) (2006) | 0.00542244 | 0.00171874 | 0.00171874 | 0.00174274 |
| [13] Zupan and Saje (2009) | 0.005429 | 0.001719 | 0.001719 | 0.001750 |
| Present work | 0.00542871 | 0.00171872 | 0.00171870 | 0.00174903 |

### *4.5 45° cantilever bending*

The fifth case is a 45 degrees cantilever beam first introduced by [18] then widely studied by several authors [13,15,17,22,23,27,28,29,30].

This case is considered in the literature as a benchmark case because all deformation modes of the beam are solicited. The beam lies in the horizontal plane with a reference configuration deformed as an arc of circle of radius $R = 100\ m$. The cross section is taken as a unit square with the following properties:

**Table 7: 45 deg bending mechanical properties**

| | |
|---|---|
| Width/height $t = h$ | 1 m |
| Young modulus $E$ | 1.e7 N/m$^2$ |
| Shear modulus $G = E/2$ | 5.e$^6$ N/m$^2$ |

Two loads of respectively 300 N and 600 N are applied at the tip. As no analytical solution exist for this specific case, Table 8 compares the results from several authors. No load increment is needed with the present shooting technique however 5 Newton iterations were needed for both load case.

The results obtained with the shooting method are in good agreement with results from the literature. This particular case involves strong nonlinear couplings between all deformation modes (shear, bending and twist). In order to achieve such precision, specific finite element formulation must be derived while the shooting method handle this couplings naturally with strong precision.

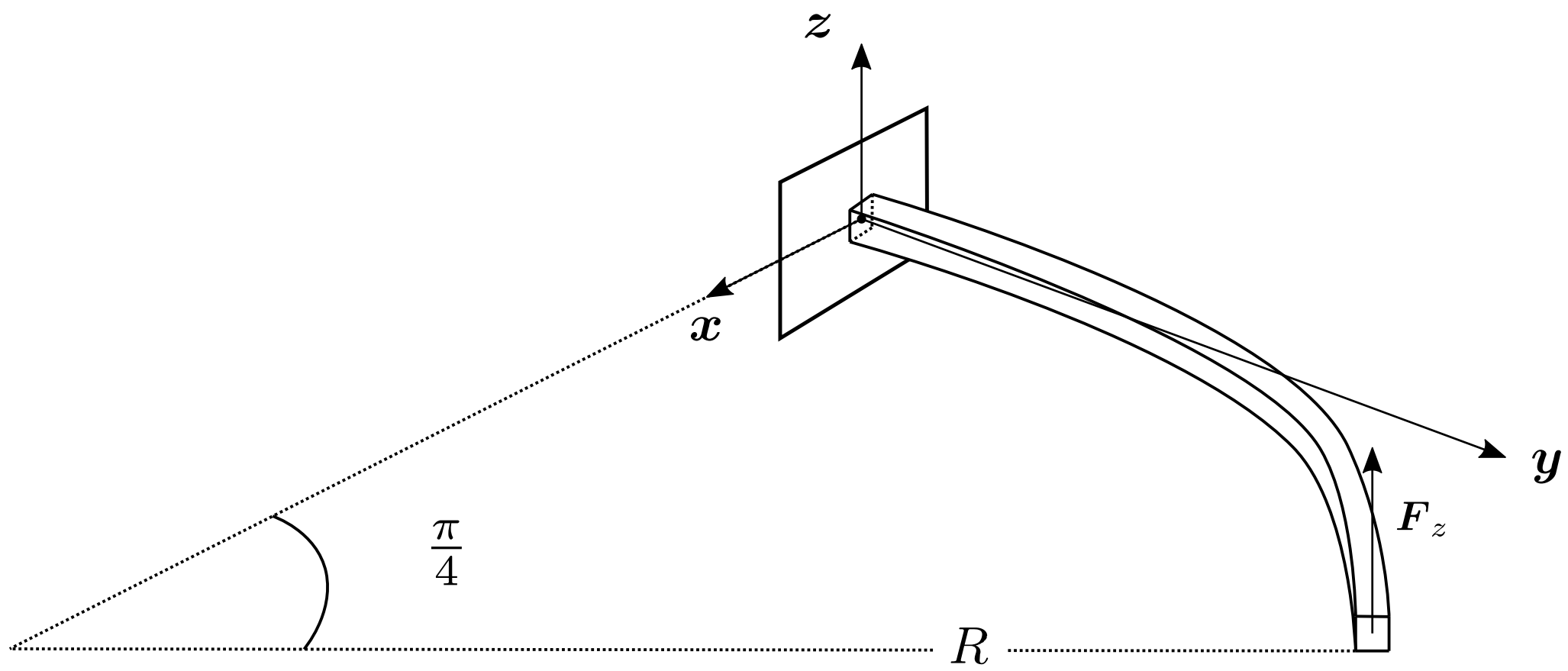

**Figure 15: 45 degrees bend cantilever**

**Table 8: 45 degrees cantilever beam submitted to F=300 N and F=600 N**

| | Tip displacement | | | | | |
|---|---|---|---|---|---|---|
| | $F = 300N$ | | | $F = 600N$ | | |
| Reference | $x$ | $y$ | $z$ | $x$ | $y$ | $z$ |
| [18] Bathe and Bolourchi (1979) | 22.5 | 59.2 | 39.5 | 15.9 | 47.2 | 53.4 |
| [15] Simo and Vu-Quoc (1986) | 22.33 | 58.84 | 40.08 | 15.79 | 47.23 | 53.37 |
| [27] Cardona and Gérardin (1988) | 22.14 | 58.64 | 40.35 | 15.55 | 47.04 | 53.50 |
| [25] Dutta (1992) | 22.20 | 58.56 | 40.46 | 15.66 | 47.02 | 53.57 |
| [28] Crivelli and Felippa (1993) | 22.31 | 58.85 | 40.08 | 15.75 | 47.25 | 53.37 |
| [16] Ibrahimbegovic, Frey, Kozar (1995) | - | - | - | 15.62 | 47.21 | 53.50 |
| [29] Smolenski (1999) | 22.19 | 58.51 | 40.25 | 15.69 | 47.01 | 53.54 |
| [23] Makinen (2007) | - | - | - | 15.62 | 47.01 | 53.50 |
| [22] Ghosh (2009) | 22.26 | 58.89 | 40.08 | 15.67 | 47.29 | 53.37 |
| [13] Zupan, Saje and Zupa (2009) | 22.14 | 58.56 | 40.47 | 15.61 | 46.89 | 53.60 |
| [30] Fan, Zhu (2016) | - | - | - | 15.68 | 47.20 | 53.45 |
| Present | 22.24 | 58.78 | 40.19 | 15.68 | 47.15 | 53.47 |

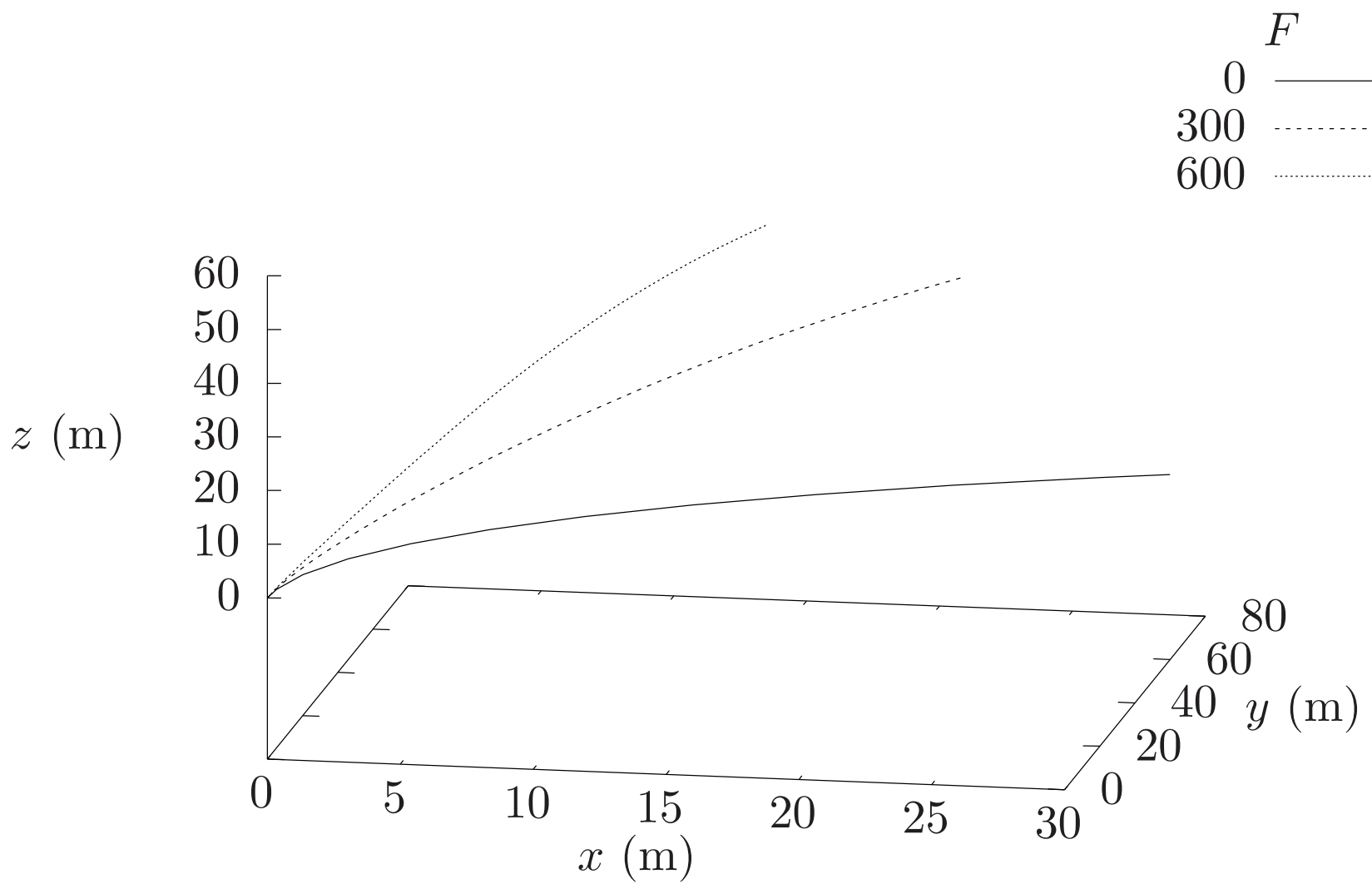

**Figure 16: 45 degrees cantilever beam submitted to F=0 N, F=300 N and F=600 N**

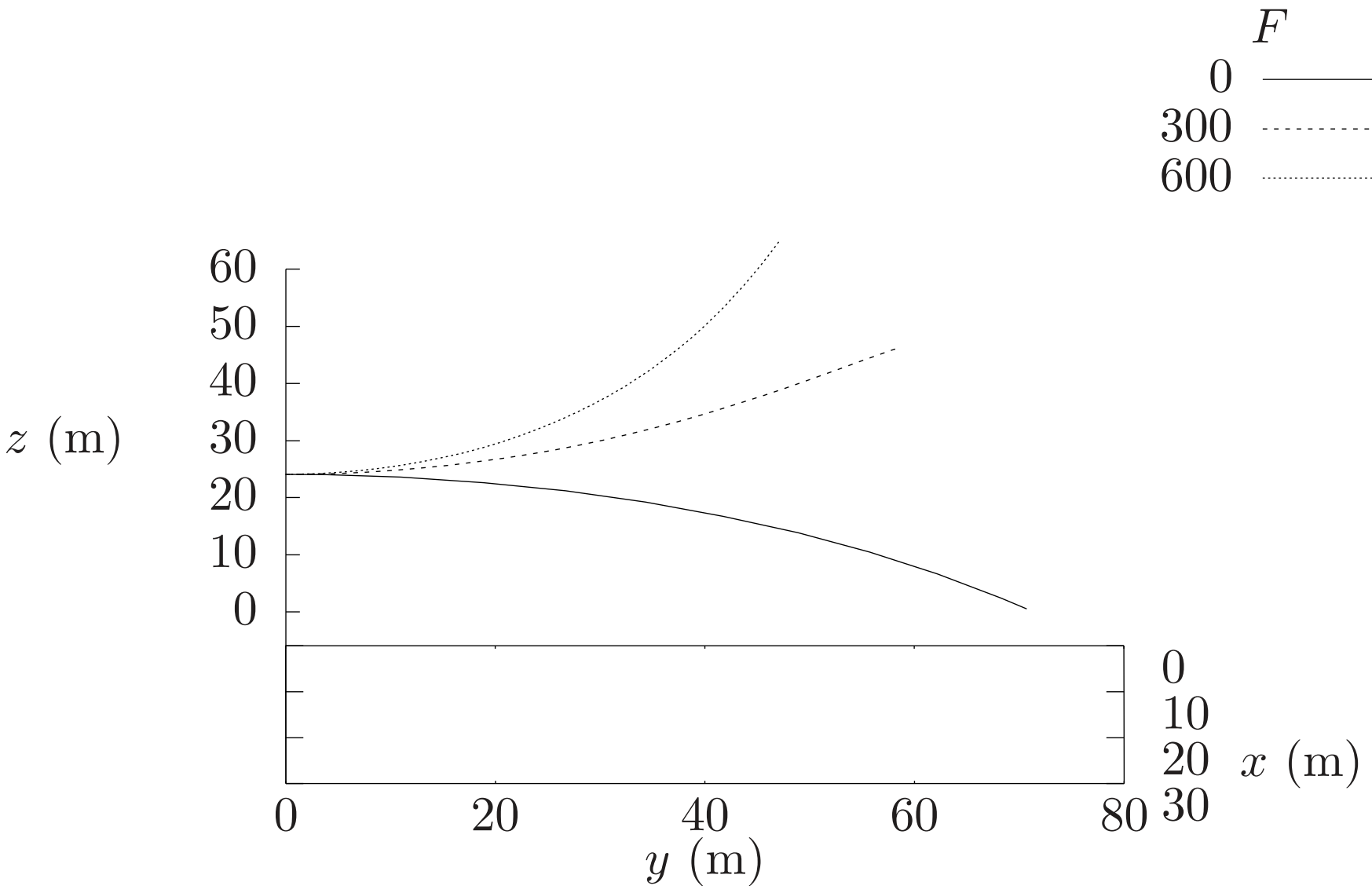

**Figure 17: 45 degrees cantilever beam submitted to F=0 N, F=300 N and F=600 N**

### *4.6 Triangularly distributed load*

The purpose of this validation case is to validate the multiple shooting method when a discontinuity is present in the external distributed load function. To do so we suppose a simple beam submitted to a symmetrically triangular distributed load as shown in Figure 18. The beam of length $L = 5\,m$ has a circular cross section of radius $r = 0.02\,m$ with a Young's modulus of $E = 2.11e11\ N/m^2$ submitted to a distributed triangular load of maximum magnitude $q = 10\ N/m$.

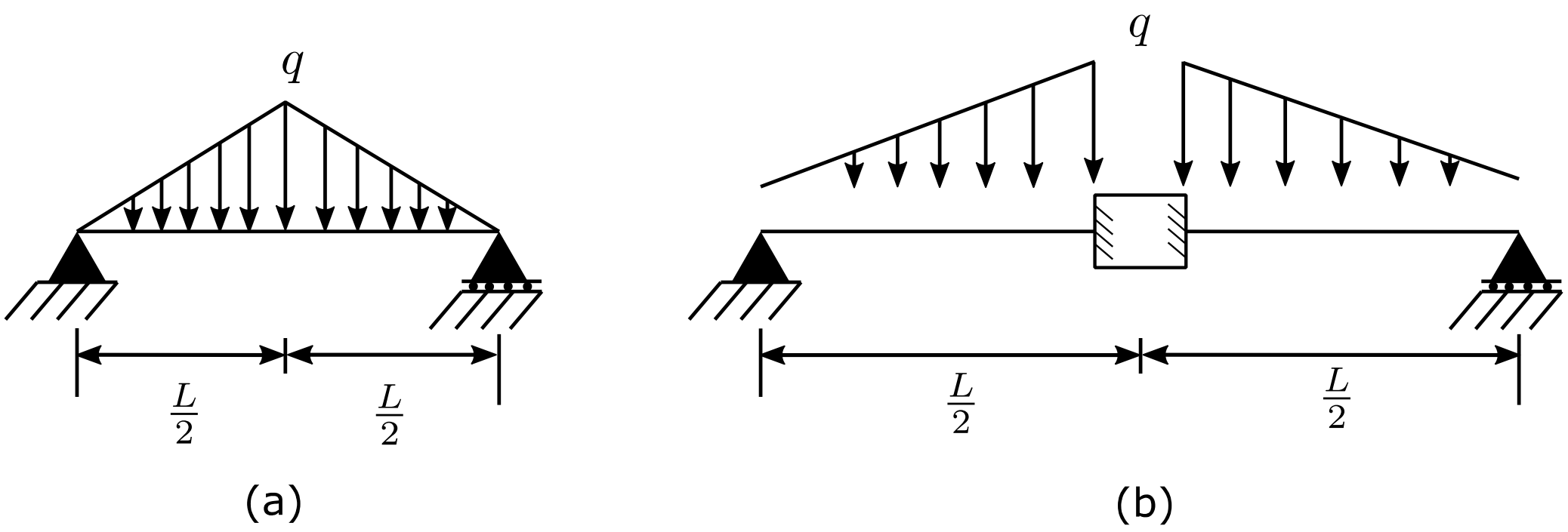

**Figure 18: triangularly distributed load (a) and its conversion in multiple shooting – rigid body formalism (b)**

Material strength formulas tell us the analytical solution of this problem is:

$$V(x) = -\frac{P}{4L}(L^2 - 4x^2) \qquad 0 \le x \le a \qquad (48)$$

$$M(x) = \frac{Px}{4L}(L^2 - 4x^2)$$

$$v(x) = -\frac{Px}{960LEI}(5L^2 - 4x^2)^2$$

$$V(x) = -\frac{P}{4L}(2x - 3L)(2x - L) \qquad a \le x \le L$$

$$M(x) = \frac{Px}{12L}(x - L)(4x^2 - 8Lx + L^2)$$

$$v(x) = \frac{P}{960LEI}(x - L)(4x^2 - 8Lx - L^2)^2$$

Figure 18 describes the beam processing in order for it to be converted in the multiple shooting method formalism with an intermediate rigid body. The initial physical segment is cut at the external force discontinuity. Each segment's end is clamped to the rigid body such that each tip and the rigid body center of gravity coincide. In this specific case, the initial physical segment curvilinear abscissa orientation is kept, saying that each IVP associated to a segment is integrated from left to right. Therefore the left segment is integrated from left boundary condition (pin) to the rigid body (clamp) whereas the second segment is integrated from the rigid body (clamp) to the right boundary condition (roller support). This integration direction is arbitrary and is not a restriction of the present method. The additional clamp boundary conditions at the rigid body are considered in the same manner as the pin and the roller boundary conditions in the shooting process (i.e. generate unknowns or constraints if they are respectively starting or ending boundary conditions). However, as mentioned previously, the rigid body has its own position and orientation degrees of freedom in order for the beam to deflect. These additional unknowns are counterbalanced by a set of additional constraints which are the static force equilibrium on the rigid body. Thus internal segment loads sum from the two clamps are equated to zero in order to generate the missing constraint. The introduction of the concept of rigid body and additional clamp boundary condition is identical from the classical multiple shooting method in this specific case.

Maximum relative error along the beam for internal shear force is 3.7303e-14, for bending moment relative error does not exceed 1.0302e-05, whereas for the maximum relative error for the deflection is 2.0287e-07. These results were obtained with a Newton's tolerance of 1.e-10. For this integrator tolerance parameter, 22 points are generated for both segments due to the symmetry of the problem.

This case demonstrate the ability of the multiple shooting method seen with rigid bodies to handle discontinuities in external distributed loading.

### 4.7 *Simple beam with two different moment of inertia*

A beam $ABCDE$ on simple support is supposed having its geometrical inertia doubled on the middle half of it. A concentrated load $P$ also acts at the midpoint $C$ of the beam. This case aims to show how the multiple shooting method with rigid bodies can model concentrated loads as well as section material discontinuities.

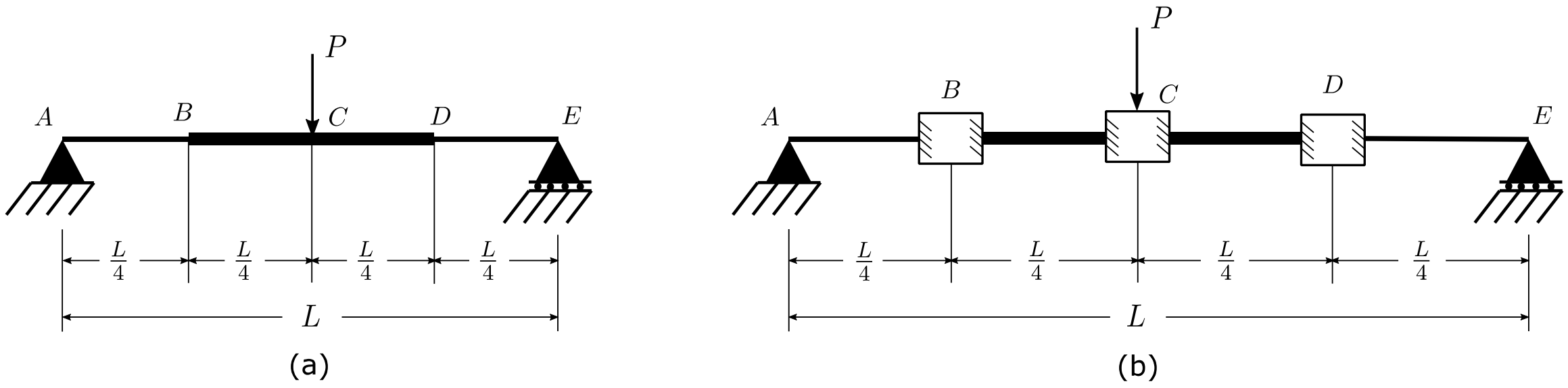

**Figure 19: simple beam with doubled geometrical inertia (a) and its conversion in multiple shooting – rigid body formalism (b)**

From material strength formulas, shear, moment and deflection for this problem are:

$$V(x) = \frac{P}{2} \qquad 0 \le x \le \frac{L}{2} \tag{49}$$

$$M(x) = \frac{Px}{2}$$

$$v(x) = -\frac{Px}{384EI}(15L^2 - 32x^2) \qquad 0 \le x \le \frac{L}{4}$$

$$v(x) = -\frac{P}{768EI}(L^3 + 24L^2x - 32x^3) \qquad \frac{L}{4} \le x \le \frac{L}{2}$$

$$V(x) = -\frac{P}{2} \qquad \frac{L}{2} \leq x \leq L$$

$$M(x) = \frac{P}{2}(L - x)$$

$$v(x) = -\frac{P}{768EI}(-7L^3 + 72L^2x - 96Lx^2 + 32x^3) \qquad \frac{L}{2} \leq x \leq \frac{3L}{4}$$

$$v(x) = -\frac{P}{384EI}(L - x)(-17L^2 + 64Lx - 32x^2) \qquad \frac{3L}{4} \leq x \leq L$$

Figure 20 to Figure 22 show the internal shear force, internal bending moment and deflection results. Numerical application supposes a beam of length $L = 10\ m$ with a circular cross section of radius $r = 0.02\ m$ with a Young's modulus of $E = 2.11e11\ N/m^2$ submitted to a concentrated load magnitude $P = 2.5\ N$.

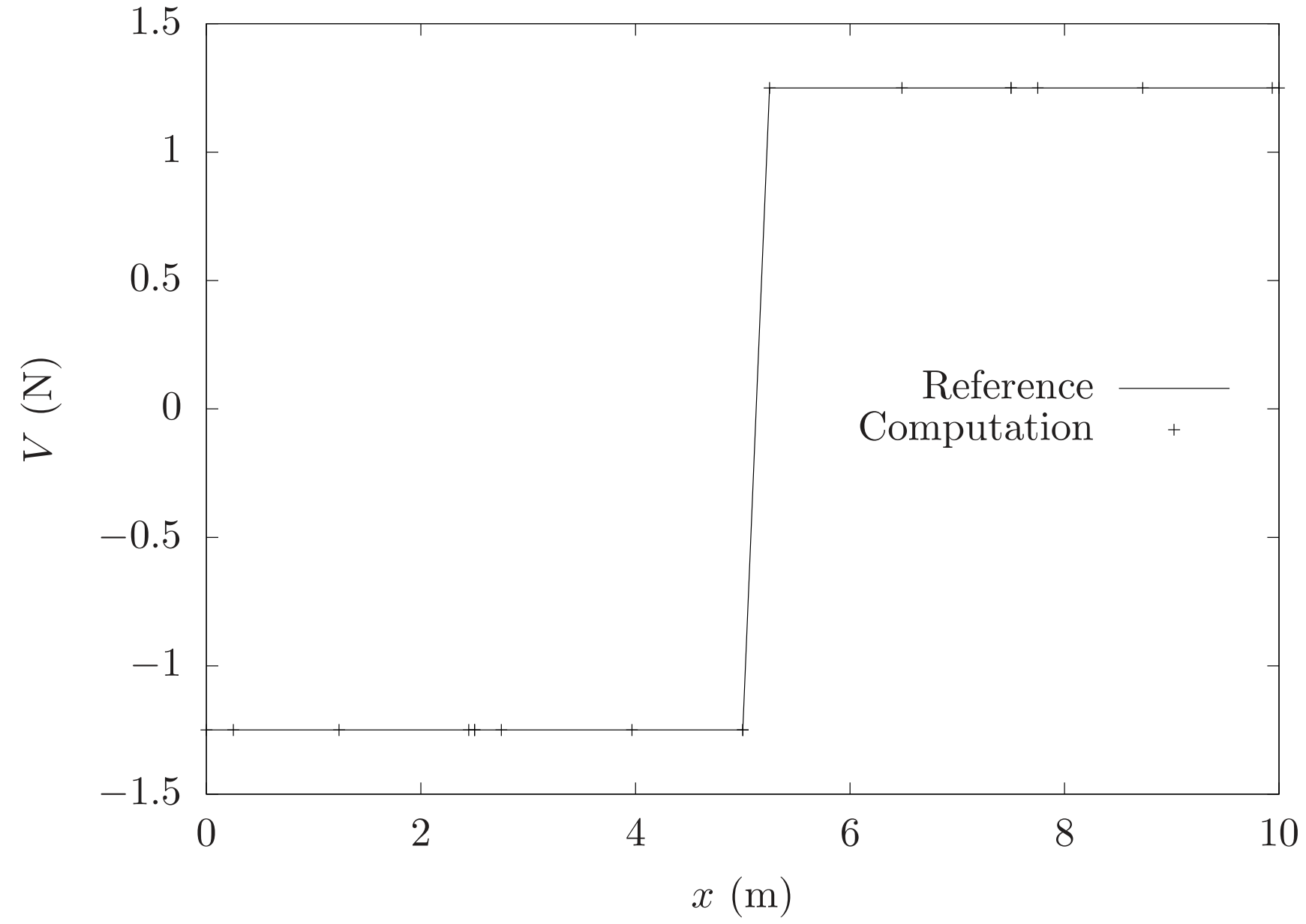

**Figure 20: simple beam with two moment of inertia – internal shear force**

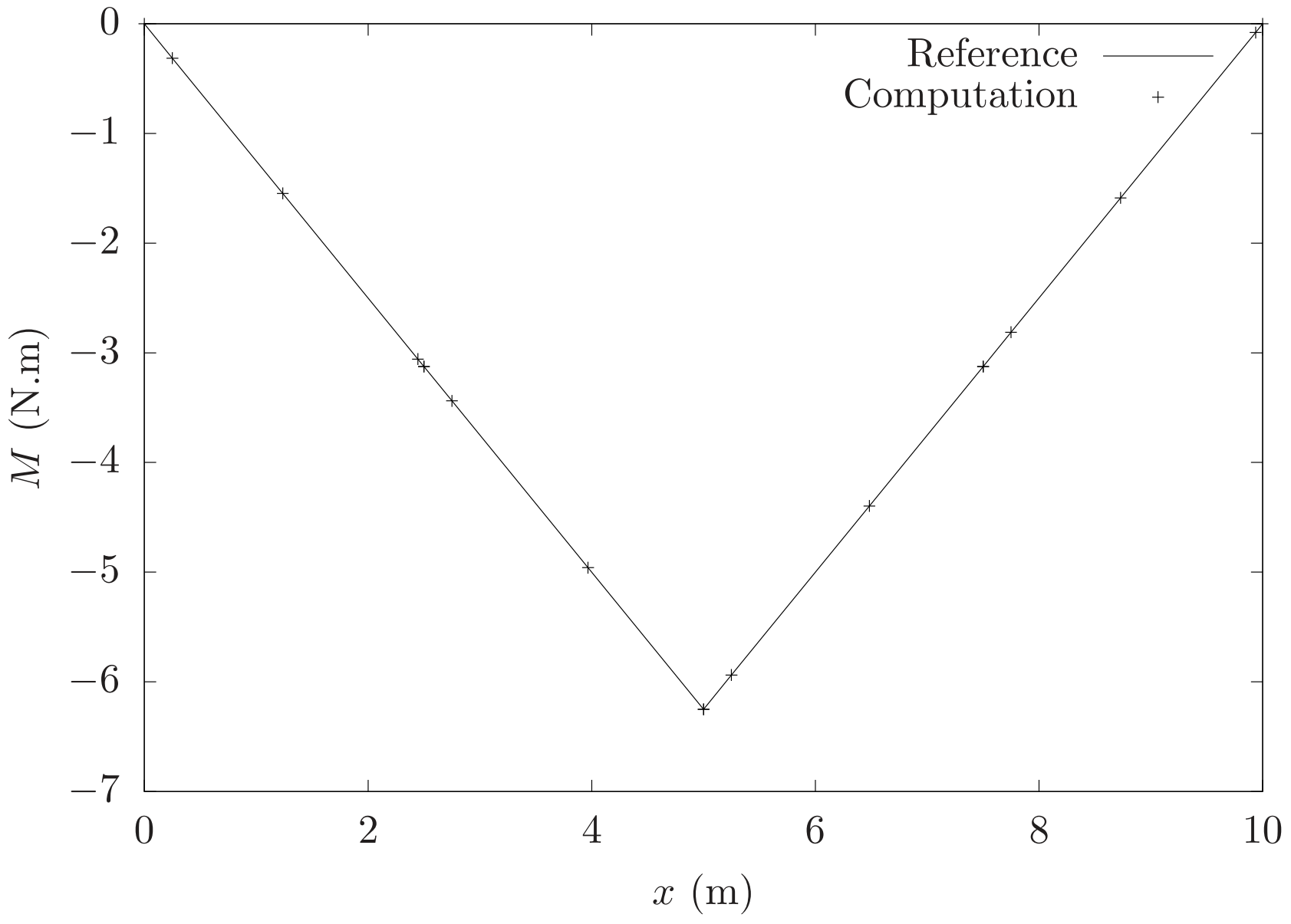

**Figure 21: simple beam with two moment of inertia – internal bending moment**

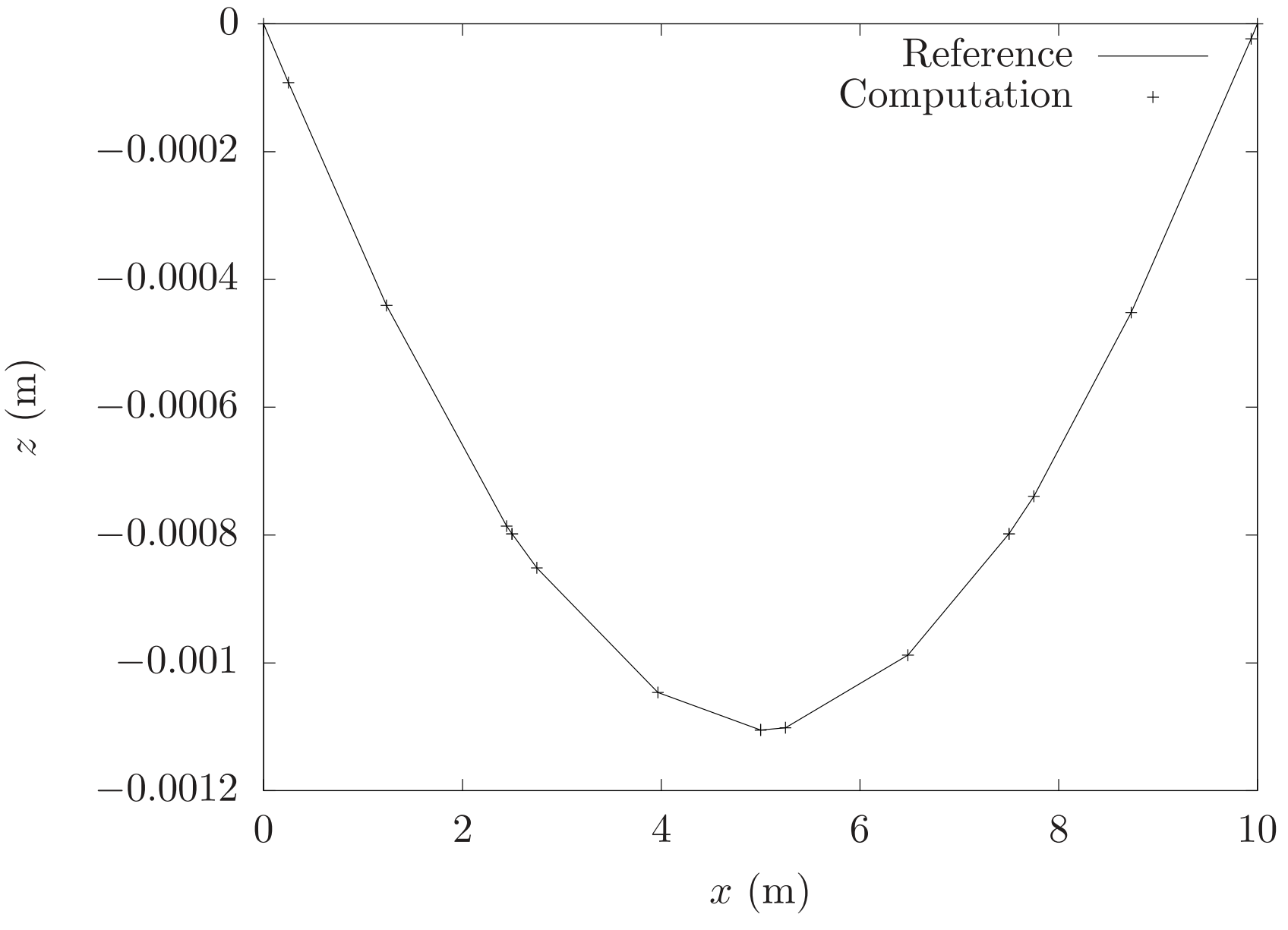

**Figure 22: simple beam with two moment of inertia – deflection**

Figure 19 show the beam representation in multiple shooting – rigid body formalism. At discontinuous section material properties, the decomposition of segments tip clamped to rigid bodies is the same as the one described in the former example of the discontinuous distributed load. However in this case, the concentrated load is directly applied to the rigid body. This leads to take into account this concentrated force contribution in the sum force constraint equation of the rigid body. These contributions come from three sources: the internal forces of segment $BC$ the internal forces of segment $CD$ and the concentrated load $P$.

Maximum relative error along the beam for internal shear force is 6.6613e-16, for bending moment relative error does not exceed 6.7815e-08, whereas for the maximum relative error for the deflection is 4.7287e-05. These results were obtained with a Newton's tolerance of 1.e-10. For this integrator tolerance parameter, 5 points are generated for the segment $AB$, 4 points for segment $BC$, 4 points for segment $CD$ and 5 points for segment $DE$ for a total of 18 points (15 if we remove duplicates at connections $B$, $C$ and $D$ of each segments).

The multiple shooting method is successfully applied here to model geometrical discontinuities in section material properties as well as concentrated loads on beams.

### *4.8 Compound beam with moment release*

This case introduce a spherical joint connecting two parts of a straight beam in order to compose with several boundary condition types linked to a rigid body. Let $ABC$ a beam composed of two segments supported by a roller support at $A$, an internal hinge (moment release) at $B$ and a clamp boundary condition at $C$ as shown in Figure 23. Segment $AB$ has a length of $a$ and segment $BC$ has a length of $b$. A concentrated load $P$ acts at a distance $2a/3$ from tip $A$, and a uniform load of intensity $q$ acts between points $B$ and $C$.

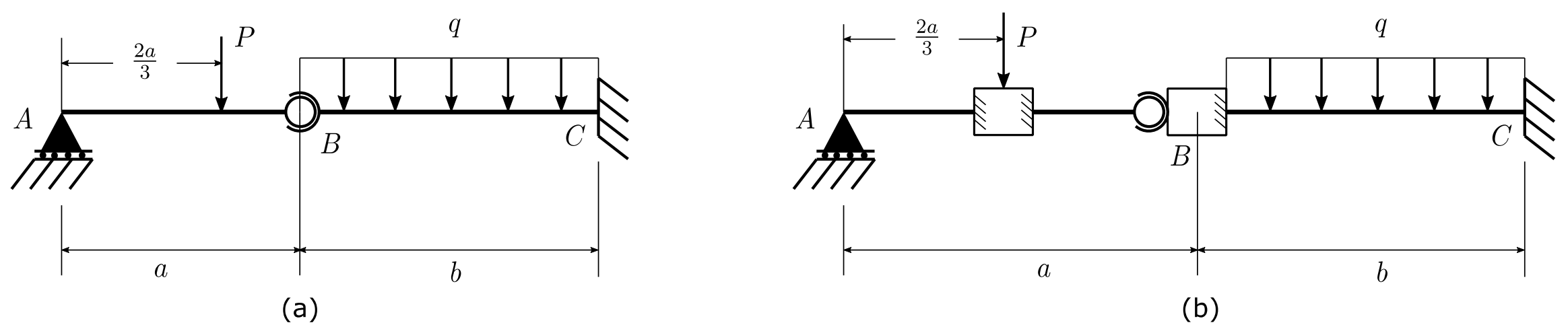

**Figure 23: compound beam (a) and its conversion in multiple shooting – rigid body formalism (b)**

From [31] deflection $\delta_B$ and angle of rotation $\theta_A$ are respectively:

$$\begin{aligned} \delta_B &= \frac{qb^4}{8EI} + \frac{2Pb^3}{9EI} \\ \theta_A &= \frac{qb^4}{8aEI} + \frac{2Pb^3}{9aEI} + \frac{4Pa^2}{81EI} \end{aligned} \tag{50}$$

The following numerical values are used for this case: $E = 2.11e11\ N/m^2$, $a = 4\ m$, $b = 6\ m$, $P = 2\ N$, $q = 1\ N/m$ and $I = \frac{\pi r^4}{4} m^4$ with $r = 0.02\ m$.

Maximum relative error for $\delta_B$ is 2.3211e-05. Maximum relative error for rotation $\theta_A$ is 2.3550e-05. Final deflection of the beam is shown in Figure 24.

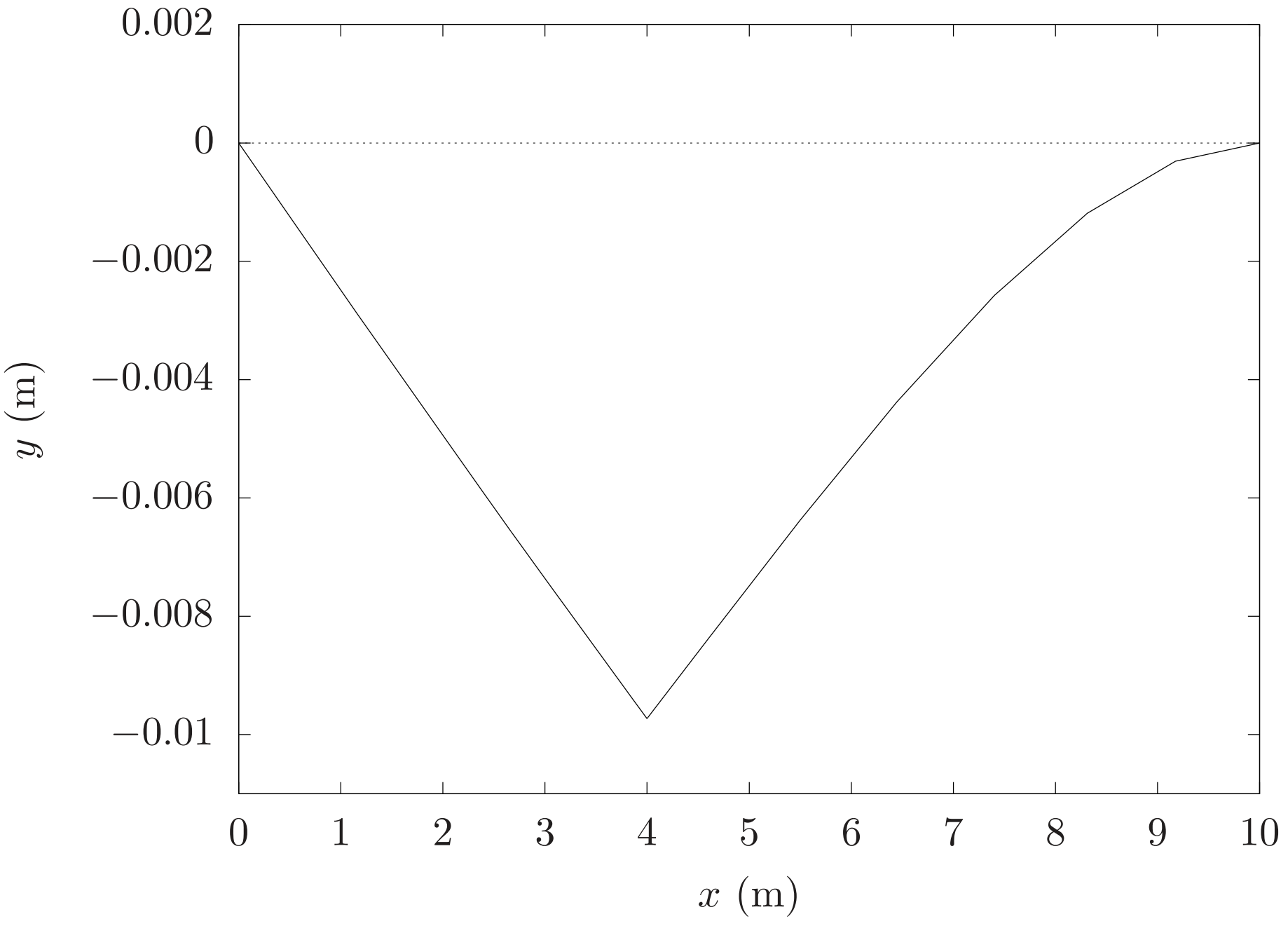

**Figure 24: compound beam – deflection**

This case enlightens the ability of the multiple shooting method combined with a rigid body description to model releases of internal kinematical degrees of freedom. Indeed one can see on Figure 23 that the tip $B$ of segment $AB$ (or indifferently tip $B$ of segment $BC$) is linked to the rigid body through a ball connection in order to model the moment release.

## *4.9 T-frame*

This final case aims to gather all elements unitary tested in previous cases. Suppose a plane T-frame $ABCD$ supported at point $A$ with a roller support and concentrated moment $M_0$ and at point $D$ with ball joint. A linearly distributed load of peak intensity $q_0$ acts on span $AB$ of the horizontal segment $ABC$. An inclined concentrated force $P$ acts at end $C$. A force with same magnitude $P$ also acts at mid-height of column $BD$.

Figure 25 show the T-frame configuration (a) as well as its conversion into multiple shooting – rigid body formalism.

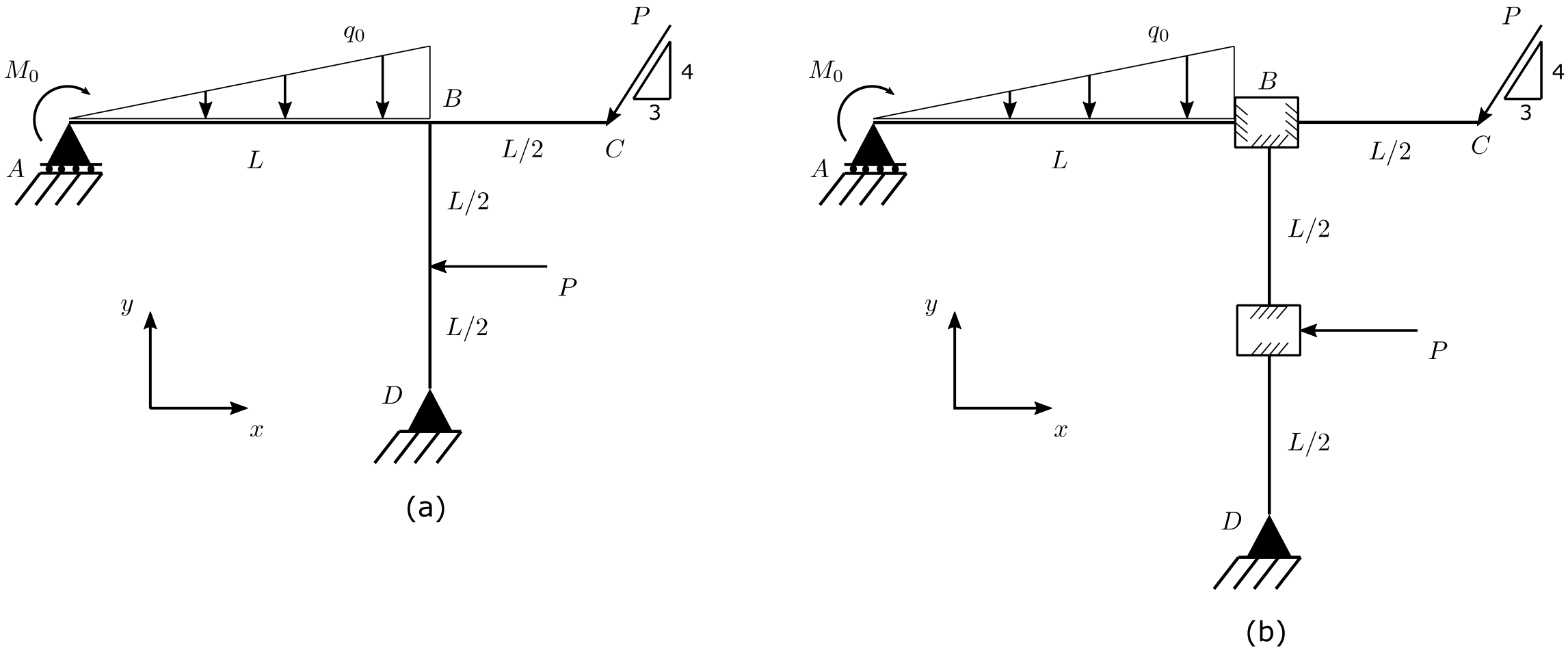

**Figure 25: T-frame (a) and its conversion in multiple shooting – rigid body formalism (b)**

From [31] internal forces and moment solution of this problem are given here below. Maximum relative errors are summarized in Table 9 for each segment.

The support reactions are:

$$D_x = \frac{8}{5}P \tag{51}$$

$$D_y = \frac{M_0}{L} + \frac{1}{3}q_0 L + \frac{1}{10}P$$

$$A_y = -\frac{M_0}{L} + \frac{1}{6}q_0 L + \frac{7}{10}P$$

On segment $AB$:

$$N(x) = 0 \tag{52}$$

$$V(x) = A_y - \frac{q_0 x^2}{2L}$$

$$M(x) = M_0 + A_y x - \frac{q_0 x^3}{6L}$$

On segment $BC$:

$$N(x) = -\frac{3}{5}P \tag{53}$$

$$V(x) = \frac{4}{5}P$$

$$M(x) = -\frac{4}{5}P\left(\frac{L}{2} - x\right)$$

On segment $DB$:

$$N(x) = -D_y \qquad 0 \le x \le L \tag{54}$$

$$V(x) = -D_x \qquad 0 \le x \le \frac{L}{2}$$

$$M(x) = -D_x x$$

$$V(x) = P - D_x \qquad \frac{L}{2} \le x \le L$$

$$M(x) = -D_x x + P\left(x - \frac{L}{2}\right)$$

**Table 9: T-frame maximum relative errors**

| Fields | $AB$ | $BC$ | $DB$ |
|---|---|---|---|
| $N$ | 4.3021e-15 | 5.9212e-16 | 8.2382e-16 |
| $V$ | 1.2357e-15 | 4.4409e-16 | 8.8818e-16 |
| $M$ | 6.7671e-16 | 7.5791e-14 | 6.6317e-14 |

Results are in very good agreement with analytical resolution from material strength formulas. The material strength solution assumes non-deformable segments in the present case. In order to have a similar model, a non-deformable constitutive law has been used here. This law expresses that strain vector $\mathbf{v}$ and curvature vector $\mathbf{u}$ are fixed and independent of internal forces $\mathbf{n}$ and moment $\mathbf{m}$ allowing no deformation. Therefore (55) and (56) are used in place of (21) and (22) in (23) and (24).

$$\mathbf{v} = \hat{\mathbf{v}}(\mathbf{n}, \mathbf{m}, s) = (0,0,1)^T \tag{55}$$

$$\mathbf{u} = \hat{\mathbf{u}}(\mathbf{n}, \mathbf{m}, s) = (0,0,0)^T \tag{56}$$

This case shows how the multiple shooting method with rigid bodies can handle geometrical discontinuities. It also allows demonstrate how the shooting method can be used together with material law other than elastic. Very good precision is achieved by doing so.

## 5. CONCLUSION

A shooting method is used to solve the TPBVP of a geometrically exact static 3D Cosserat rod. The presented formalism is applicable to single shooting as well as multiple shooting. The static equations are kept fully nonlinear in a material description. This paper focused specifically on static approach. The use of the shooting method for dynamic beam equations is not straightforward. Indeed, dynamic beam equations are a TPBVP governed by Partial Differential Equations (PDE) while the shooting method is a tool to solve TPBVP governed by Ordinary Differential Equations (ODE). However, the shooting method has been used for time domain simulation where the time domain derivative is discretized through a backward difference Euler scheme [32]. Spatial derivatives are kept continuous and the discretized terms are passed on the right term of equilibrium equations. This lead to solve static equations augmented with inertial terms. For frequency analyses, the shooting method has also been extensively used to find periodic solutions (natural modes) of dynamic systems [33,34].

In the presented formalism, a generic continuous constitutive law is introduced, with applications regarding linear elastic or non-deformable behavior. This demonstrates the generality of the method in the modeling of material laws. Cross section properties and spatial coordinates as well as internal forces and strains are kept continuous with regards to arc length. All sets of equations, including the contribution of external distributed loads are written in 3D.

Cosserat rods kinematics undergoing large displacements and rotations kinematics are derived. This work proposed a kinematic singularity free orientation parameterization based on MRP which has been investigated in the shooting method process. The switching relation to their shadow set SMRP is reminded and an algorithm to bypass orientation singularity while performing the ODE integration of the shooting method is given.

The boundary conditions to solve the TPBVP are expressed as functions of the initial and final state variables. A review of all applicable beam boundary condition types and their inclusion in the shooting method process has been made. In the present case of separated boundary conditions, the constraint functions are derived from the admissible statics of mechanical linkages. Several boundary condition types combinations have been studied and the shooting method has shown strong results accuracy for each one of them.

It has been shown that the use of a RK45 adaptive step algorithm to solve the IVP brings a limited number of integration points while keeping very good accuracy with regards to all beams fields whether it be geometrical fields or internal loads fields. Thus good computation time is reached.

A root finding based on Newton's algorithm has been used in the present paper in order to solve the constraint function $\boldsymbol{C}$. The Newton approach is appropriated to find the roots of the constraint vector function, the convergence is quadratic and several solutions can be found. However Newton's algorithm is known for its lack of robustness when the starting point is too far from a solution. This limitation turns out to be true especially in the case of stiff problems, i.e. $\nabla\boldsymbol{C}$ is badly conditioned. Such cases have been encountered in specific ball-ball beams configurations as well as in near buckling studies where the IVP itself is badly conditioned. Because the multiple shooting method consists in dividing integration intervals in smaller intervals which tends to linearize the constraint function $\boldsymbol{C}$ of the initial problem over each subinterval, a solution may still be reachable when the single shooting method fails. But this convergence problem is a key point to be overcome in order for the shooting method to be robust enough. If the remark on conditioning difficulty is relevant for the study of beams, the strings problem is less sensible to this convergence issue as shown in [7]. Compared to the finite elements methods, the shooting method benefits of an almost analytical precision and a stepsize free convergence. Two future developments are considered: firstly formulate the root finding problem as a minimization process with the help of a functional in order to use stronger optimization algorithm than Newton's approach; secondly determine a functional that leads to better conditioning than the current formalism.

Finally, the classical single and multiple shooting method principles have been reminded. A special emphasis on the multiple shooting method formalism is done by interpreting it as a single shooting method generalization through a re-parameterization of the subdomains in which lies the local curvilinear abscissa.

The proposed boundary conditions formalism together with the introduction of the concept of rigid bodies led to the development of a new shooting method which enhances the capabilities of the classical multiple shooting method. This allows the modeling of assembled beams subjected to discontinuities coming from several sources in the 1D static equations: external distributed or concentrated forces, material cross sections, and geometrical discontinuities. By doing so, the multiple shooting method with rigid bodies naturally found its place in a multi-body framework.